\RequirePackage{fix-cm}
\documentclass[smallextended]{svjour3}       
\smartqed  
\usepackage{graphicx}
\usepackage{dsfont}
\usepackage{mathrsfs}
\usepackage{amssymb}
\usepackage{enumerate}
\usepackage{amsmath}
\usepackage{caption}
\usepackage{subfig}
\usepackage{appendix}

\newtheorem{thm}{Theorem}
\newtheorem{dfn}[thm]{Definition}
\newtheorem{rem}[thm]{Remark}
\newtheorem{prop}[thm]{Proposition}
\newtheorem{lem}[thm]{Lemma}
\newtheorem{notation}[thm]{Notation}
\newtheorem{ass}[thm]{Assumption}

\newcommand{\one}{\mathds{1}}
\newcommand{\calF}{\mathscr{F}}
\newcommand{\BP}{\mathbb{P}}
\newcommand{\BQ}{\mathbb{Q}}
\newcommand{\RR}{\mathbb{R}}
\newcommand{\BE}{\mathbb{E}}
\newcommand{\BN}{\mathbb{N}}
\newcommand{\BS}{\mathbb{S}}
\newcommand{\HH}{\mathbb{H}}
\newcommand{\BM}{\mathbb{M}}
\begin{document}

\title{XVA Valuation under Market Illiquidity
}


\author{Weijie Pang        \and
         Stephan Sturm  
}


\institute{W. Pang \at
              Department of Mathematics and Statistics, McMaster University\\
              \email{pangw6@mcmaster.ca}           
           \and
           S. Sturm \at
              Department of Mathematical Science, Worcester Polytechnic Institute\\
              \email{ssturm@wpi.edu}
}

\date{Received: date / Accepted: date}

\maketitle

\begin{abstract}
Before the 2008 financial crisis, most research in financial mathematics focused on pricing options without considering the effects of counterparties' defaults, illiquidity problems, and the role of the sale and repurchase agreement (Repo) market. Recently, models were proposed to address this by computing a total valuation adjustment (XVA) of derivatives; however without considering a potential crisis in the market. In this article, we include a possible crisis by using an alternating renewal process to describe the switching between a normal financial regime and a financial crisis. We develop a framework to price the XVA of a European claim in this state-dependent situation. The price is characterized as a solution to a backward stochastic differential equation (BSDE), and we prove the existence and uniqueness of this solution. In a numerical study based on a deep learning algorithm for BSDEs, we compare the effect of different parameters on the valuation of the XVA.
\end{abstract}

\section{Introduction}
\label{chap:intro}

The 20$^{\rm th}$ century witnessed the birth and development of financial mathematics, but it had been varying dramatically since the 2008 financial crisis.
While one decisive cause of this crisis is the unhealthy mortgage market. The percentage of sub-prime mortgage, compared with all mortgages, doubled in 2004 and continued to increase. During June 2004 and June 2006, the Fed fund rate was increased from 1.00\% to 5.25\%, and stayed at 5.25\% until the Federal Reserve started decrease the rate in September 2007. Considering the hazardous mortgage market at that time, numerous households experienced grievous financial pressure, because of shortages of adequately financial capacity. Some sold their house and escape from their mortgage luckily, others had to default on their mortgage loans. In both situations, those houses returned U.S. real estate market. Because the supply of houses pumped up, prices of real estate dropped sharply, which consisted with the Law of Supply and Demand. Meanwhile many investment companies lost dramatically in their loans in the unhealthy mortgage market and their investments in the real estate. As a result, companies experienced serious financial suffer as well. Several financial companies even went bankrupt, such as Lehman Brothers. The 2008 sub-prime financial crisis resulted out from this cause. 

According to this situation, many research attempt to take the effect of defaults from counterparty and debt into account in the replication framework. In order to consider defaults from a institution and its counterparty in a pricing model of an option, the bebt valuation adjustment (DVA) and the credit valuation adjustment (CVA)  become popular. For the spread between interest rates from and to a bank, researcher add a funding valuation adjustment (FVA) to their original framework. Many distinct valuation adjustments come out after the 2008 financial crisis.  Since Basel regulations have required to include default risk and cost of collateralization strategies in risk measurement (\cite{basel2010basel}), increasing people work on the total valuation adjustment (XVA), as a combination of CVA, DVA, FVA and several other valuation adjustments.

Research on FVA had been starting before the 2008 financial crisis. Before the crisis,
\cite{cvitanic1993hedging} introduce the setting of different interest rates for borrowing and lending in a stochastic control problem for hedging. To price European call and put options, \cite{korn1995contingent} assume that the borrowing interest ratea is higher than the lending one. Based on a similar assumptions of asymmetric interest rates, \cite{el1997backward} study super-hedging by nonlinear backward stochastic differential equations. \cite{laughton2012defence} emphasize the necessity of a funding valuation adjustment (FVA) in an incomplete market. \cite{siadat2017net} discuss the hedging, collateral optimization and reverse stress testing with funding cost and funding benefit adjustment. In addition to a funding account,
\cite{piterbarg2010funding} include a sale and repurchase agreement (Repo) market to calculates the adjustment for non-collateralised derivatives. 

Besides the FVA, there are numerious meaning results in the topics of CVA and DVA.  \cite{bielecki2008arbitrage} and \cite{crepey2010counterparty} study the valuation and hedging of credit default swap (CDS) including counterparty credit risk. \cite{crepey2013counterparty} extend a total valuation adjustment model to several derivitives, using a Markovian pre-default backward stochastic differential equation. Besides counterpary credit risk, \cite{brigo2013counterparty} take funding cost and collateral service cost into account, and derive a risk-neutral pricing formula.
Initially, the pricing model only included a default from a counterparty (unilaterally model), which lead to the asymmetric pricing result. To solve this problem, it is necessary to consider defaults from an company and its counterparty together. From then on, people introduced bilateral models and build several framework.
\cite{burgard2011balance,burgard2011partial} generalize Piterbarg's model to include bilateral credit risk. In theoretical field, \cite{nie2014fair} prove the existence of a fair bilateral price. \cite{bielecki2013credit} introduce a general semimartingale market framework for an arbitrage-free valuation.  \cite{bichuch2015arbitrage,bichuch2016arbitrage,bichuch2018arbitrage} include financing from a sale and repurchase agreement (Repo) market, and construct a backward stochastic differential equation representation of European call and put option prices with bilateral credit risk, asymmetric funding, and collateral rates. Recently, \cite{bichuch2018robust} extend the valuation of the XVA with considering the uncertainty bond rates. 
\cite{bielecki2018arbitrage} study the nonlinear arbitrage-free pricing of derivatives, considering differential funding cost, collaterlization, counterparty credit risk, and capital requirements.

The 2008 financial crisis histronically change pricing models, the study of the sale and repurchase agreement (Repo) markets had been altering as well. 
A Sale and Repurchase Agreement (Repo) is the sale of a security combined with an agreement to repurchase the same security at a specied price at the end of the contract. \cite{hordahl2008developments} point out that the Repo market froze during the financial crisis and compare distinct performances among U.S., Euro and U.K. Repo market and deduce the reasons of the differences.  Focus on the U.S. Repo market, \cite{gorton2012securitized} conclude that the bilateral Repo market became one important channel of the spreading for the sub-prime financial crisis. Contrarily, \cite{krishnamurthy2014sizing} belive that the funding of securitized assets has enough source, although Repo activities involving private-label securitized assets sizing down. Furthermore, \cite{gorton2010regulating,thiemann2015regulation} analyze regulations to normalize the Repo market. \cite{copeland2014repo} construct an infinite-horizon equilibrium model to investigate the fragility of Repo market, according to different microstructures and financial corporates. However, few research include the effect of Repo markets in a financial crisis in their study of the XVA.




In order to seperate a financial crisis from a normal financial crisis, many different indicators have been proposed. \cite{hollo2012ciss} introduce a Composite Indicator of Systemic Stress (CISS), which puts more weight on the stress shared by several markets at the same time.  \cite{whaley2009understanding} argues that the CBOE's Market Volatility Index (VIX), introduced in \cite{whaley1993derivatives}, is a good investor fear gauge of the expected return volatility of the S\&P 500 index over the next 30 days. \cite{adrian2011covar} introduce the value at risk of financial institutions conditional on other institutions being in distress (abbreviated CoVaR) as a new measure for systemic risk. The difference between the three month London Interbank Offered Rate (LIBOR) and the government's interest rate for a three-month period (called the Ted spread) changed alot during the financial crisis. As a result, \cite{heider2009liquidity} and \cite{acharya2011model} argue that the Ted spread is a good indicator for the liquidity and counterparty risk in the interbank system. \cite{mancini2013liquidity,coffey2009capital,gorton2012securitized,gorton2010regulating} use the Ted spread to measure the capital constraints in a secured lending system. In addition, \cite{boudt2013funding} confirm the existence of a two-regime Ted spread over the period between 2006 and 2011, describing stable and unstable situation by the analysis of historical data. However, these difference regimes have not been considered in pricing derivatives so far.

Overall, few research includes all of the bilateral credit risk, asymmetric interest rates, andfunding and market illiquidity, causing by the frozen of Repo markets during a financial crisis. Therefore, we are interested in the pricing of options while considering defaults of an investor and its counterparty, funding illiquidity problem and switching between different financial regimes.
In other words, this paper focuses on the total valuation adjustment for one European option with investor's and its counterparty's defaults, different financial regimes, and funding liquidity problems. Its organization is as follows. In Section \ref{sec:repo}, we review several topics about the Repo market. In Section \ref{sec:status}, we apply an alternating renewal process to describe the switching between financial regimes. After the discussion of several financial accounts in Section \ref{sec:replicating}, we create a hedging portfolio for European options, construct a BSDE to evaluate the arbitrage-free price of a European option and prove the existence and uniqueness of the solution in Section \ref{sec:pricing}. In Section \ref{sec:xva}, we construct a BSDE of the XVA and derive a corresponding reduced BSDEs to a smaller filtration. In Section \ref{sec:pricing_simu}, we estimate the parameters of the alternating renewal process by the Ted spread historical data and analyze the sensitivity of the XVA with respect to the financial states, volatilities and  funding rates.

\section{Sale and Repurchase Agreement (Repo) Market}\label{sec:repo}
In this section, we review several necessary topics about the Sale and Repurchase Agreement (Repo) market at first. Then, we analyze its different performances during a calm financial time and a financial crisis period as well as its influence to a stock market.

\subsection{Structure of Repo markets}
A Sale and Repurchase Agreement (Repo) is a sale of a security combined with an agreement to repurchase the same security at a specified price at the end of the contract. Over the last 40 years, the size of the Repo market increased dramatically. From 2002 to 2007, its capital size even doubled. More details in \cite{gorton2010regulating,gorton2012securitized}.

A contract in the Repo market specifies two transactions. At the beginning, one party sells a specific security to the counterparty at a given price. At the end of the contract, the party repurchases the same security from its counterparty at the agreed price, which was decided by two parties during the contract's negotiation. Here, the specific security can be seen as a collateral in a collateralized borrowing transaction. The collateral provider is also a cash receiver, and the collateral receiver is also a cash lender. In this terminology, the above transactions can also be explained in another way. At the initial time, the cash provider (collateral receiver) lends $m$ dollar to its counterparty (cash receiver, collateral provider). At the same time, the collateral provider (cash receiver) gives a security as collateral to the cash provider (collateral receiver). At the maturity time, the cash receiver (collateral provider) returns the $m+r$ dollars to the cash provider (collateral receiver). At the same time, the collateral receiver (cash provider) returns the collateral to the collateral provider (cash receiver). 
\begin{figure}[h]
	\centering
	\includegraphics[width=0.6\textwidth]{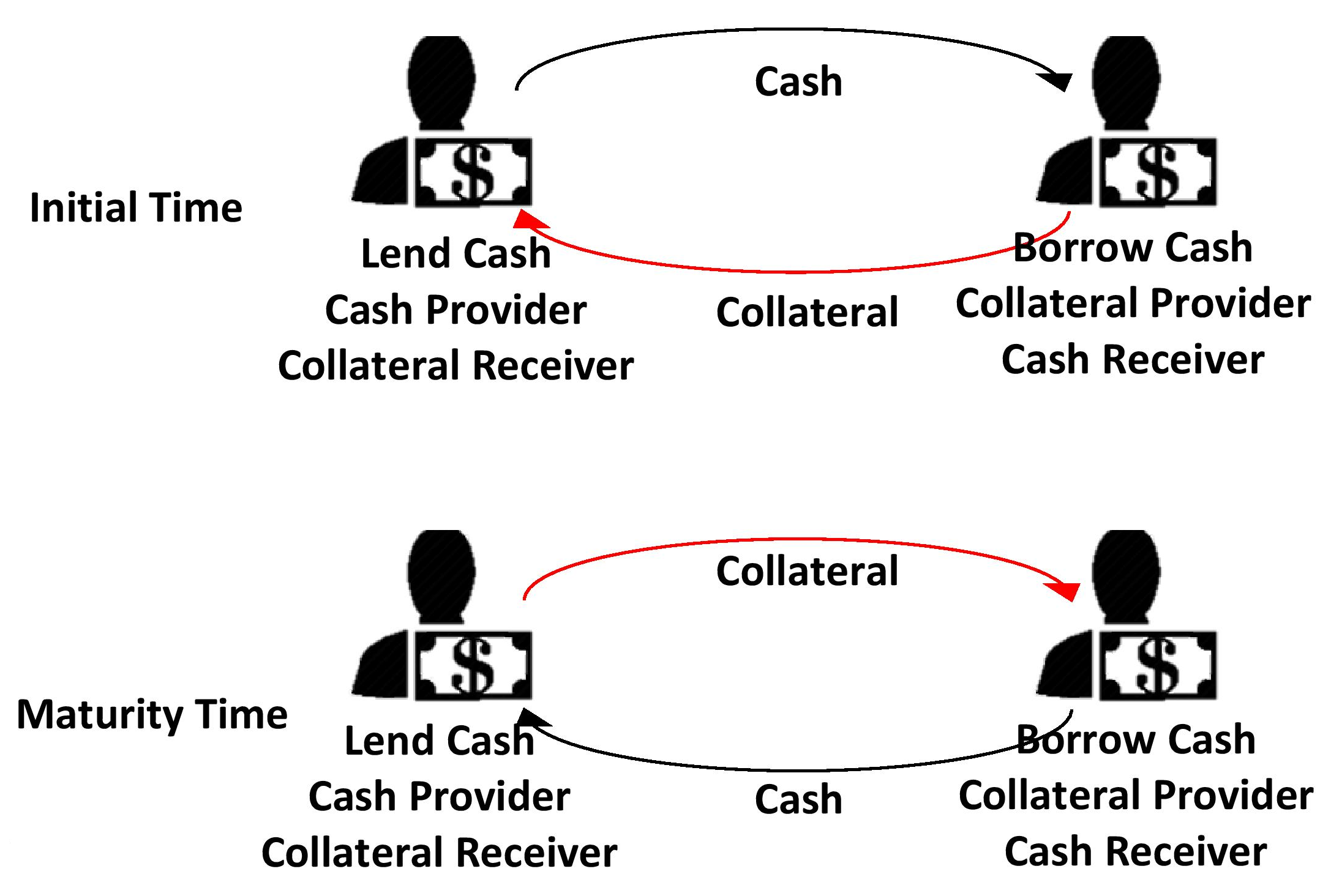}
	\caption{Transactions in Repo Market.}
\end{figure}

In this transaction, a relative difference between the two cash flows $\frac{(m + r - m)}{m} = \frac{r}{m}$
is called a Repo rate. Usually, the market value of the collateral is larger than the cash transaction. For example, when borrowing $m$ dollars, one needs to provide a collateral with a market price as $m+h$ dollars. The relative difference between the market price of the collateral and the cash lent is called the haircut, $\frac{m+h-m}{m} = \frac{h}{m}$. Based on different confidences played in the collateral, the haircut varies from 0.5\% to over 8\% in a calm financial period. In the U.S. Repo market, a group of safe collaterals is called the general collateral, it includes, i.e. 10 years U.S. treasury bonds.

Based on the numbers of participants and the party to holding the collaterals, Repo markets have three classical types -- Bilateral Repo, Triparty Repo and Hold-in-custody Repo. The Bilateral Repo has been already introduced in the beginning of this section. For the Triparty Repo, there is an agent between two parties in this Repo transactions. The agent helps to prepare the contract and also hold the collateral in its balance sheet during the lifetime of the contract. A Hold-in-custody Repo can be a Bilateral Repo or a Triparty Repo, but the collateral is held on the balance sheet of the collateral provider during the lifetime of the contract.

This market is a significant source of collateralized borrowing of cash and borrowing of any specific securities. When a Repo contract is used to borrow cash, it is called a cash driven Repo activity.  On the other hand, many companies use the Repo market as a source to borrow a specific security to meet their liquidity requirements, which is called a security driven Repo activity. To attract collateral providers of some special securities, the cash lender may provide very low Repo rate, even a negative value. This unusual Repo rate is call a special Repo rate.

\subsection{Performance in Sub-prime Financial Crisis}

During the 2008  financial crisis, the United State bilateral Repo market froze mostly. At the beginning of the crisis, the declining price of securities caused panic of these securities in investors. Because most traders lost their confidence on these securities, Repo transactions involving these securities were rejected. Contrarily, using the general collaterals were welcome for all collateral recivers. Nevertheless, traders holding the general collateral were unwilling to lend them to other traders, especially after some general collateral receivers defaulted and rejected to return those collaterals. As a result, the bilateral Repo trading based on the general collateral also froze. Overall, we assume that there was no contracts at bilateral Repo markets during a financial crisis. More details in \cite{gorton2012securitized}.

The frozen bilateral Repo market affected the stock market during a financial crisis. Since the Repo market is an important source of illiquid securities, many traders conduct their short sells of stocks by borrowing the stocks at a Repo market and then selling them in a stock market. In fact, more than 90\% of the contracts in the bilateral Repo market are short-term contracts, normally for only one day. Based on the frozen situation of the U.S. bilateral Repo markets, the source to borrow a specific stock is extremely insufficient.
Meanwhile, the Securities and Exchange Commission (SEC) banned the short-selling of 733 companies' stocks on September 19$^{th}$, and then extended the ban to more 190 companies. The purpose of this ban is to maintain or restore fair and orderly securities markets. Thus, most short stock trades ceased during the financial crisis. More information about funding liquidity and market liquidity can be found at \cite{brunnermeier2008market}.


\section{The Status Process}\label{sec:status}

As we mentioned in Section \ref{sec:repo}, both the Repo market and stock market have distinct performance and rules between a calm financial period and a miserable period. In order to describe the switching between a normal financial regime and a financial crisis for these two markets, we need to study a jump stochastic process, called as alternating renewal process. 

\subsection{Definition}
An alternating renewal process is a non-homogeneous Poisson process without an independent increments property. This process is well known in engineering field, where it is used to describe a service period and a shutdown period of one machine or one system.  Thus, it is also known as On-Off process.

\begin{figure}[!h]
	\centering
	\includegraphics[width=0.6\textwidth]{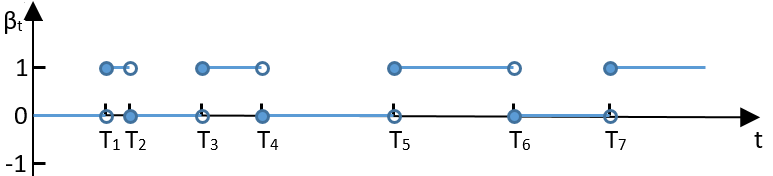}
	\caption{One path of the process $\beta$.}
	\label{fig:betapath}
\end{figure}

An alternating renewal process, denoted as $\beta$, is a stochastic process switching between $0$ and $1$, as shown in Figure \ref{fig:betapath}. The mathematical defition is as follow.
\begin{dfn}\label{def:beta}
	An alternating renewal process at time $t >0$ is defined as 
	$
	\beta_t = \sum_{i=1}^\infty (-1)^{i+1}\one_{\{ T_i \leq t \}}, 
	$
	where odd inter-arrival times $T_{2n+1} - T_{2n}$ follow an exponential distribution with a constant parameter $\lambda_U > 0$, and even inter-arrival times $T_{2n} - T_{2n-1}$ follow an exponential distribution with a constant parameter $\lambda_V > 0$.
\end{dfn}

\begin{rem}
	Corresponding to Definition \ref{def:beta}, an alternating renewal process $\beta$ is a stochastic process without independent increments property when $\lambda_U \neq \lambda_V$.
\end{rem}

In this paper, we apply an alternating renewal process to describe those switchings between a normal financial status and a financial crisis status. When $\beta_t = 0$, financial markets are in a calm period. When $\beta_t = 1$, financial markets are in a financial crisis period. 
Based on the definition of the alternating renewal processes, a holding time of a calm status follows an exponential distribution with the parameter $\lambda_U$ and a holding time of a financial crisis follows an exponential distribution with the parameter $\lambda_V$.

\begin{rem}
	In this paper, we apply two-regime setting for the financial market. This can be extend to a more general multi-regimes. For example, a three status of financial market: a calm regime, a stress regime and a crisis period. 
\end{rem} 

For the alternating renewal process, there are many interesting topics about its properties and distribution. Besides the non-independent increment property, we only introduce one important theorem about its intensity, which proof can be find in Appendix \ref{chap:poisson}. Define a filtration $(\calF_t^\beta)_{t\geq 0}$, where $\calF_t^\beta= \sigma(\beta_s: s\leq t)$.
\begin{thm}\label{thm:betaintensity}
	For an alternating renewal process $\beta$, there exists a finite variation stochastic process $\Lambda^\beta$ such that $ \tilde{\beta}_t := \beta_t - \Lambda^\beta_t = \beta_t - \int_0^t \lambda^\beta_s ds$, $\tilde{\beta}$ is a martingale with respect to the filtration $(\calF_t^\beta)_{t \geq 0}$. 
\end{thm}

Based on the definition and property of the alternating renewal process, we also define a non decresing jump counting process, denoted as $J_t$.
\begin{dfn}
	For the alternating renewal process $\beta$ with parameters $\lambda_U$ and $\lambda_V$, we define a jump counting process $J$ at time $t > 0$ as
	\begin{equation}\label{jumpdef}
	J_t = \sum_{s \leq t}\one_{\{\beta_s - \beta_{s-} \neq 0\}}(s).
	\end{equation}
\end{dfn} 
Similar to the process $\beta$, we have the following remark and proposition.
\begin{rem}
	The jump counting process $J$ is a stochastic process without independent increments property when $\lambda_U \neq \lambda_V$.
\end{rem}

\begin{prop}\label{Jsi}
	The jump counting process $J$ is a square integrable semimartingale with respect to the filtration $(\calF_t^\beta)_{t \geq 0}$.
\end{prop}
\proof{Proof}
Proof in Appendix \ref{chap:poisson}
\endproof

\begin{prop}\label{Jtilde}
	For the jump counting process $J$, there exists a finite variation stochastic process $\Lambda^J$ such that $\tilde{J}_t := J_t - \Lambda^J_t = J_t - \int_0^t \lambda_s^J ds, \tilde{J}$ is a square integrable martingale with respect to the filtration $(\calF_t^\beta)_{t \geq 0}$.
\end{prop}
\proof{Proof}
Proof in Appendix \ref{chap:poisson}
\endproof

\section{Replicating Portfolio}\label{sec:replicating}

In order to price an European option, we need to construct a replicating portfolio. Besides a classical stock account, we include accounts of two risky bonds to consider credit risks from the debt and the counterparty. In order to contain liquidity problems about the funding and the market, we take the funding account and Repo account into our portfolio. In this section, we review these assets in a filtered probability space $(\Omega, \calF, (\calF_t)_{t\geq 0}, \BP)$, which is rich enough to provide all necessary information, containing information about stock prices, default of risky bonds and switching between a normal financial regime and a financial crisis. Here, the probability $\BP$ is the physical probability measure. Moreover, all interest rates are asymmetric, which means the rates of borrowing and lending are distinct.

\subsection{Repo Account}

We already analyze the structure of the Repo market and its performance during the 2008 financial crisis. In this paper, an investor apply the Repo market as a source to  borrow or lend cash and stocks. Considering a setting of the asymmetric interest rates, we assume that Repo rates are different constants for cash lenders and cash borrowers. For cash lenders, they receive constant interest rate $r_r^+$ in Repo markets. For cash borrowers, they pay constant interest rate $r_r^-$, and implement long positions. 

Let $\psi_t^r$ be a number of shares of a Repo account, then the Repo rate is 
$$
r_r(\psi^r) = r_r^- \one_{ \{ \psi^r < 0 \}} + r_r^+ \one_{\{ \psi^r > 0\}}.
$$
In a calm financial regime, we represent the Repo account as $B^{r_r^-}$ and $B^{r_r^+}$ for a borrowing and a lending, respectively. Overall, the dynamics of one Repo account is $dB_t^{r_r^{\pm}} = r_r^{\pm} B_t^{r_r^\pm} dt$. So, the value of one Repo account at time $t$ is 
$$
\psi_t^r B_t^{r_r} = \psi_t^rB_t^{r_r} (\psi^r) = \psi_t^r \exp\Big( \int_0^t r_r(\psi^r_s) ds \Big).
$$

However, the U.S. bilateral Repo market froze during the 2008 financial crisis, as we mentioned in Section \ref{sec:repo}. In order to distinguish the values of a Repo account during a calm financial regime and a financial crisis, we contain the alternating renewal process $\beta$ in Section \ref{sec:status} in the description of the value of a Repo account. In this paper, we assume all Repo contracts are overnight contracts. With a frozen bilateral Repo market during a financial crisis, the value of the Repo account is zero during a crisis ($\beta = 1$). Thus, we represent the Repo account as
$$
(1-\beta_t) \psi_t^rB_t^{r_r}  = (1-\beta_t)\psi_t^r\exp\left(\int_0^t r_r (\psi^r_s) ds\right),
$$
The dynamics of the Repo account is
$$
(1-\beta_t)\psi_t^rdB_t^{r_r^\pm} = (1-\beta_t) r_r^\pm \psi_t^r  B_t^{r_r^\pm} dt .
$$

\subsection{Funding Account}

Besides the Repo market, the investor can also borrow and lend cash through the treasury desk. In order to replicate a European option, we also include the funding account into our replicating portfolio. In practice, the interest rates for cash borrowers and cash lenders are different. In this paper, we also apply this aymmetric assumption for the funding interest rates. For cash lenders, they receive a constant interest rate $r_f^+$ from the treasury desk. For cash borrowers, they pay a constant interest rate $r_f^-$ to funding desks. Let $\xi_t^f$ \index{$\xi_t^f$} be a number of shares in the funding account at time t. So the funding interest rate is represented as 
$$
r_f := r_f(\xi)
= r_f^- \one_{ \{  \xi < 0 \}} + r_f^+ \one_{ \{  \xi >0  \}}.
$$

Similarly with the Repo account, the funding account is
$$
\xi_t^f B_t^{r_f} := \xi_t^f B_t^{r_f} (\xi^f) = \xi_t^f \exp\left(\int_0^t r_f(\xi_s^f)ds\right),
$$
with the dynamics
$
dB_t^{r_f^\pm} = r_f^\pm B_t^{r_f^\pm} dt.
$

\subsection{Stock Security}

In this paper, the price of the stock follows a classical geometric Brownian motion in a calm financial period. We assume that $W^\BP$ is a standard Brownian motion under the physical measure $\BP$ and the inital value is $S_0$, then the dynamic of the price is  
$$
dS_t = \mu S_t dt + \sigma S_t dW_t^\BP,
$$
where $\mu$ is a constant drift rate and $\sigma$ is a constant volatility. When denoting the number of shares of the stock as $\xi_t$, the value of a stock account is $\xi_t S_t$ in a normal finanical period.

Contrarily to a calm regime, the short selling of stocks are banned, as a result of the combination of a frozen Repo market and short-selling ban during a financial suffering regime. In other words, the value of a stock account cannot be a negative number during a financial crisis ($\beta = 1$). Summarizing these different performances of a stock accounts, we describe the value of one stock account as
$$
(1- \beta_t \one_{\{\xi_t < 0 \}}) \xi_t S_t.
$$
Here the term $(1- \beta_t \one_{\{\xi_t < 0 \}})$ is the adjustment, corresponding to the frozen short trades in a financial crisis. The dynamic of the stock account is 
$$
(1- \beta_t \one_{\{\xi_t < 0 \}}) \xi_t dS_t = (1- \beta_t \one_{\{\xi_t < 0 \}}) \xi_t ( \mu S_t dt + \sigma S_t dW_t^\BP).
$$

\subsection{Risky Bond Securities}
We use the defaults of an investor's risky bond and its counterparty's risky bond to replicate the defaults of the investor and its counterparty, respectively. 
By abbreviating an investor as ``I" and its counterparty as ``C", we represent
the default times of these two sides as two stopping times $\tau_I$ and $\tau_C$, respectively. We assume that default times follow exponential distributions with constant intensities $h_i^\BP, i \in \{I, C\}$. Then the default indicator processes of these two risky bonds are described as
$$
H^i_t = \one_{ \{  \tau_i \leq t \}},  \quad \forall t \geq 0 ,\qquad  i\in \{I, C\}.
$$ 

The default stopping times follow an exponential distribution with parameters $r^i + h_i, i \in \{ I,C \}$, respectively. So, the two risky bond prices $P^I_t$ and $P^C_t$, underwritten by the investor and the counterparty, have the following dynamics
$$
dP_t^i = (r^i + h_i^\BP) P_t^i dt - P_{t-}^i dH_t^i,\quad P_0^i = \exp(-(r^i+ h_i^\BP)T),
$$
where $r^i + h_i^\BP, i \in \{I, C\}$ are constant return rates, respectively.

Therefore, for these two default indicator processes $H_t^t, i \in \{I, C\}$ with constant parameters $h_i^\BP, i \in \{I, C\}$, we have 
$
\varpi_t^{i,\BP} = H_t^i - \int_0^t (1-H_u^i) h_i^\BP du
$
is a $(\calF_t)_{t \geq 0}$ martingale under the measure $\BP$.

\subsection{Collateral Account}
The collateral account is to protect one party from the counterparty's default. Similarly with the funding account and the Repo account, the value of the collateral process $C:= (C_t: t \geq 0)$ also has a pair of asymmetric constant collateral rates. For collateral receivers ($C_t < 0$), they pay a constant interest rate $r_c^-$. For collateral providers ($C_t > 0$), they receive a constant interest rate $r_c^+$.  We represent the collateral cash account as $B^{r_c^\pm}$ by representing collateral interest rates as
$
r_c(c) =  r_c^- \one_{ \{  c < 0 \}} + r_c^+ \one_{ \{  0 < c  \}}.
$
The collateral cash account is 
$$
B_t^{r_c} := B_t^{r_c} (C) = \exp\left(\int_0^t r_c(C_s) ds \right).
$$
Based on this definition, its dynamics is
$
dB_t^{r_c^\pm} = r_c^\pm B_t^{r_c^\pm} dt.
$
Denote a number of shares of the collateral account $B_t^{r_c}$ by $\psi_t^c$, we have 
\begin{equation}\label{eq:collateral}
\psi_t^c B_t^{r_c} = -C_t.
\end{equation}
This means that a collateral receiver ($C_t < 0$) should purchase $\psi_t^c > 0$ shares of the collateral account, and vice versa. Assume that $\hat{V}$ be a third party valuation of the European claim, i.e. the Black-Scholes price, the collateral account is assigned different collateralization levels based on the safe levels of the counterparty. This is modeled as 
$$
C_t: = \alpha \hat{V},
$$
where the collateralization level $\alpha$ is a constant between 0 and 1. When the counterparty is very reliable, for example the U.S. government, the collateralization level $\alpha = 0$. When the counterparty has low credit rating, the collateralization level could be $\alpha = 1$. More detials in \cite{bichuch2018arbitrage}.

\subsection{Wealth Process}

After reviewing all financial assets in previous section, we build a replicating portfolio when taking an investor and its counterparty's defaults, liquidity problems in a Repo market and a stock market, and asymmetric interest rates into account. In general, a classical replicating portfolio only two accounts, a underlying stock account and a funding account. However, our replicating portfolio has four other accounts, the Repo account, two risky bonds accounts and the collateral account. The representations and dynamics of these six accounts are already reviewed in previous section. Our investing strategy is a six dimensional vector, denoted as  $\varphi := (\xi_t, \xi_t^f, \xi_t^I, \xi_t^C, \psi_t^r, \psi_t^c; t \geq 0)$. Here, $\xi_t$ represents shares of stocks, $\xi_t^f$ denotes shares of the funding account, $\xi_t^i, i \in \{I, C\}$ are shares of risky bonds, underwritten by the investor an its counterparty, respectively, $\psi_t^r$ describes shares of the Repo account, and $\psi_t^c$ symbolizes shares of the collateral account. By descriptions of all assets in Section \ref{sec:replicating}, we describe our replicating portfolio as
\begin{equation}\label{wealtheq}
V_t(\varphi) = (1- \beta_t \one_{\{ \xi_t < 0 \}})\xi_t S_t + \xi_t^I P_t^I + \xi_t^C P_t^C + \xi_t^f B_t^{r_f} + (1 - \beta_t) \psi_t^r B_t^{r_r}  - \psi_t^c B_t^{r_c}.
\end{equation}

The dynamics of this wealth process $V_t$ is
$$
dV_t(\varphi) = (1- \beta_t \one_{\{ \xi_t < 0 \}})\xi_t dS_t + \xi^I_t dP_t^I + \xi_t^C dP_t^C + \xi_t^f dB_t^{r_f} + (1 - \beta_t) \psi_t^r dB_t^{r_r} - \psi_t^c dB_t^{r_c}.
$$

\begin{dfn}\index{Self-financing}
	An investment strategy is self-financing if for any $t \in [0,T]$, the following identity
	{\small
		\begin{equation*}
		V_t(\varphi) = V_0(\varphi) + \int_0^t(1- \beta_t \one_{\{ \xi_t < 0 \}})\xi_t d S_t + \int_0^t\xi_t^I dP_t^I + \int_0^t \xi_t^C dP_t^C + \int_0^t \xi_t^f dB_t^{r_f} + \int_0^t(1 - \beta_t) \psi_t^r dB_t^{r_r}  - \int_0^t \psi_t^c dB_t^{r_c},
		\end{equation*}}
	where $V_0(\varphi)$ is the initial capital.
\end{dfn}

\subsection{Closeout Valuation}

In general, the closeout value of a classical replicating portfolio is the value of one option at its maturity. However, our portfolio has more complicate closeout valuation when we consider defaults of the investor and its counterparty. In this paper, we do not consider the fire sales of the collateral account at the default times $\tau^I$ and $\tau^C$. Let $\tau = \tau^I \wedge \tau^C \wedge T $ be the maturity time of our portfolio. Assume $0 \leq L_I, L_C \leq 1$ be loss rates of investors and counterparties at their default time. Let $\hat{V}$ be a third party valuation of the hedging portfolio, i.e. a classical Black-Scholes valuation of one option. At default times, the value of the claim is represent as  $Y:= \hat{V}_\tau - C_\tau = (1-\alpha) \hat{V}_\tau$. Therefore, the closeout value is $\hat{V} - L_IY^+$ when the investor defaults, and it is $\hat{V} - L_CY^-$ when the counterparty defaults, where $(\cdot)^+ = \max (0, (\cdot))$ and $(\cdot)^- = \max(0, -(\cdot))$. Overall, we have a closeout valuation of our replicating portfolio at time $\tau$ as
\begin{equation*}
\begin{split}
\theta(\tau, \hat{V}) :=  &\one_{\{ \tau^I < \tau^C \}} \theta_I(\hat{V}_\tau) + \one_{ \{ \tau^C< \tau^I \} } \theta_C (\hat{V}_\tau)\\
=&\hat{V}_\tau + \one_{\{\tau^C < \tau^I  \}} L_C Y^- - \one_{\{ \tau^I<\tau^C \}} L_I Y^+,\\
\end{split}
\end{equation*}
where $\theta_I(v) : = v - L_I((1-\alpha) v)^+ , \theta_C(v) : = v + L_C((1-\alpha )v)^-$. Here, the term $\one_{ \{\tau^C < \tau^I\} } L_CY^-$ is the credit valuation adjustment term after collateral mitigation and the term $\one_{ \{ \tau^I<\tau^C \} } L_IY^+$ is the debit valuation adjustment term.

\section{Arbitrage-Free Pricing}\label{sec:pricing}

In this section, we evaluate the claim by the principle of No-Arbitrage. Firstly, we convert the dynamics of all financial assets in the real measure $\BP$ to a valuation measure. Then we discuss necessary and sufficient conditions of the No-Arbitrage and construct backward stochastic differential equations (BSDEs) of the wealth process to evaluate the European options on this valuation measure. At end of this section, we prove the existence and uniqueness of the solutions to our BSDEs of the wealth process.

\subsection{Valuation Measure}
In Section \ref{sec:replicating}, we already give the descriptions of all financial assets in a real probability $\BP$. But in order to apply the principle of No-Arbitrage, we need to convert the stochastic processes of stock and risky bonds account to processes with a consistent drift rate. Because all interest rates are asymmetric in this paper, we cannot use the normal risk-free interest rate as the discount rate. Here we assume the investor can choose a discount rate for this valuation model, which is denoted as $r_D$. By the Radon--Nikodym derivative, we define the valuation measure $\BQ$ with respect to a discount rate $r_D$ as
\begin{equation*}
\begin{split}
\frac{d\BQ}{d\BP}\Big|_{\calF_t} =& \exp\left( \frac{r_D - \mu }{\sigma } W^\BP_t - \frac{(r_D - \mu )^2}{2\sigma^2} t \right) \\
&\left(1+ \frac{r^I - r_D}{h^\BP_I} \right)^{H_t^I}\exp((r_D-r^I)t) \left( 1 + \frac{r^C - r_D}{h^\BP_C} \right)^{H^C_t} \exp((r_D- r^C)t).
\end{split}
\end{equation*}

We denote $\mu_I = r^I + h_I^\BP$ and $\mu_C = r^C + h_C^\BP$, which are return rates of risky bonds, underwritten by the investor and counterparty, respectively. The above equation becomes
\begin{equation*}
\begin{split}
\frac{d\BQ}{d\BP}\Big|_{\calF_t} =& \exp\left( \frac{r_D - \mu }{\sigma } W^\BP_t - \frac{(r_D - \mu )^2}{2\sigma^2} t \right) \left(\frac{\mu^I - r_D}{h^\BP_I} \right)^{H_t^I}\\
&\exp((r_D-\mu_I +h_I^\BP)t) \left( \frac{\mu^C - r_D}{h^\BP_C} \right)^{H^C_t} \exp((r_D-\mu^C + h_C^\BP)t).
\end{split}
\end{equation*}

Under this valuation measure $\BQ$, the dynamics of three risky assets are 
\begin{equation}\label{dynq}
\begin{split}
dS_t = r_D S_t dt + \sigma S_t dW^\BQ_t,\\
dP_t^I  = r_D P^I_t dt - P^I_{t-} d\varpi_t^{I,\BQ},\\
dP_t^C = r_D P^C_t dt - P^C_{t-} d\varpi_t^{C,\BQ},
\end{split}
\end{equation}
where 
$$
W_t^\BQ = W^\BP_t - \frac{r_D - \mu}{\sigma }t
$$
is a Brownian Motion under $\BQ$ and 
$$
\varpi_t^{i,\BQ} = \varpi_t^{i,\BP} + \int_0^t (1-H_u^i)(h_i^\BP - h_i^\BQ)du, i \in \{I,C\}
$$
are $((\calF_t)_{t \geq 0}, \BQ)$ martingales. Here $h^\BQ_i = \mu_i - r_D \geq 0$ are default intensities of default indicator processes under $\BQ$.

\subsection{Arbitrage-free Assumptions}
\begin{dfn}[Arbitrage]
	The market admits an investor's arbitrage, if there exists a investment strategy $\varphi = (\xi_t, \xi_t^I, \xi_t^C, \xi_t^f, \psi_t^r, \psi_t^c; t \geq 0)$ such that 
	$$
	\BP[V_t(\varphi, x) \geq \exp(r_f^+ t) x] = 1, \quad \BP[V_t(\varphi, x) > \exp(r_f^+ t) x] > 0,
	$$
	for a given initial capital $x \geq 0$ and a corresponding wealth process $V(\varphi, x)$.
\end{dfn}

\begin{dfn}[Arbitrage-free Financial Markets]
	If a financial market does not admit an investor's arbitrage for any $x \geq 0$, 
	the market is arbitrage free from the investor's perspective.
\end{dfn}

In Section \ref{sec:replicating}, we already review all financial assets. In order to ban the opportunity of an arbitrage in the financial market, we need the following assumptions.
\begin{ass}\label{noarbinec}Necessary Assumptions of Arbitrage-free Financial Markets
	\begin{enumerate}
		\item $r_f^+ \leq r_f^-,$
		\item $r_f^+ \vee r_D < \mu_I \wedge \mu_C,$
		\item $(1-\beta_t)r_r^+ \leq (1-\beta_t)r_f^-$ (i.e. $r_r^+ \leq r_f^-$ in a normal financial status).
	\end{enumerate}
\end{ass}

\begin{rem} 
	These three assumptions are necessary to exclude an arbitrage potentiality.
	\begin{enumerate}
		\item If $r_f^+ > r_f^-$, one can borrow cash from the funding desk at a funding rate $r_f^-$, and then lend it to the funding desk at the funding rate $r_f^+$. There is a positive arbitrage profit of $r_f^+ - r_f^- > 0$ multiplies the amount of cash.
		
		\item If $r_f^+ > \mu_I$ (or $r_f^+ > \mu_C$), one can short sell investor's (or a counterparty's) risky bonds with an expected return rate $\mu_I$ (or $\mu_C$), and then lend the money to the funding desk, earning an arbitrage profit $r_f^+ - \mu_I > 0$ (or $r_f^+ - \mu_C > 0$) multiplies the shares of the bonds. 
		
		If $r_D > \mu_I$ (or $r_D > \mu_C$) and the investor can trade by the interest rate $r_D$, the discussion is similar to the cases $r_f^+ > \mu_I$ (or $r_f^+ > \mu_C$). If the investor cannot trade by $r_D$, then the investor's default intensity is $h_I^\BQ = \mu_I-r_D < 0$ (or $h_C^\BQ = \mu_C-r_D < 0$) under the discount measure $\BQ$, which is not realistic.
		
		\item In a normal financial status ($\beta_t = 0$), if $r_r^+ > r_f^-$, one can borrow cash from the funding desk at the funding rate $r_f^-$ and lend it to the Repo market at the Repo rate $r_r^+$, earning a positive arbitrage profit $r_r^+ - r_f^- > 0 $ multiplies the amount of cash. In a financial crisis status ($\beta_t = 1$), the inequality holds trivially.
	\end{enumerate}
\end{rem}

\begin{prop}\label{noarbiass}
	Suppose that Assumption \ref{noarbinec} holds. A financial market is arbitrage-free if 
	$$
	(1-\beta_t)r_r^+ \leq (1-\beta_t)r_f^+ \leq (1-\beta_t)r_r^- ,
	$$
	which means $r_r^+ \leq r_f^+ \leq r_r^-$ in the normal financial status.
\end{prop}

\proof{Proof}
In a calm financial market, this arbitrage-free condition is already proved in \cite{bichuch2018arbitrage}. During a financial crisis, the result is a direct result from a multiplier $1-\beta_t$ and the proof in \cite{bichuch2018arbitrage}.	
\endproof

\begin{rem}
	If an investor knows the information of the financial status ($\beta$), this financial market is still arbitrage-free, based on the previous proposition.
\end{rem}

\subsection{Dynamics of The Wealth Process}

In Section \ref{sec:repo}, we already discuss the relationship between the Repo account and the stock account. In order to avoid ``naked" short selling of a stock, short-selling traders need to ``borrow" the stock from aRepo market befor they sell it in the stock market. In this paper, we ignore the haircut in the Repo trading. Then the Repo account and the short selling of stocks have the following equation.
$$
\psi_t B_t^{r_r} = - \xi_t S_t, \quad t > 0,
$$
when $\xi_t < 0$. Moreover, when investors need to borrow cash to purchase stocks, they prefer Repo market than the funding desk in general, considering the efficiency and resilience of the Repo market. So the above equation holds whenever $\xi_t < 0$ or $\xi_t \geq 0$ during a calm finanical period. To sum up, we have the following assumption in a calm financial market.

\begin{ass}\label{ass:repostock}
	In a normal financial status, the stock account is financed by the Repo market, which is represented as
	$$
	(1-\beta_t)\psi_t^rB_t^{r_r} = - (1-\beta_t)\xi_tS_t.
	$$
\end{ass}

Because borrowing in the Repo market is a collateralized borrowing trade, the borrowing Repo rate $r^-_r$ is less than the uncolleralized funding rate $r_f^-$ in general.

Contrary to a calm period, we already discuss the performance of the bilateral Repo market and stock market during a financial crisis in Section \ref{sec:repo} and Section \ref{sec:replicating}. Since the bilateral Repo market freeze during a financial crisis, an investor has to borrow money from the funding desk to buy stocks. Considering both of the normal financial status and the financial crisis status, we summarize the relationship of all asset accounts as 
\begin{equation}\label{equalaccount}
(1-\beta_t)\psi_t^r B_t^{r_r} + \beta_t \xi_t^f B_t^{r_f} = \beta_t V_t + \beta_t\psi_t^c B_t^{r_c} -  (1-\beta_t \one_{\{\xi_t < 0\}})\xi_t S_t - \beta_t \xi_t^I P_t^I - \beta_t \xi_t^C P^C_t.
\end{equation}

By Equations \eqref{wealtheq}, \eqref{dynq}, and the self-financing property of the replicating portfolio, the dynamics of $V_t$ under the valuation measure $\BQ$ with $r_D$ is
\begin{equation*}
\begin{split}
dV_t = 
&(1-\beta_t \one_{\{\xi_t < 0\}}) \xi_t dS_t + \xi_t^I dP_t^I + \xi_t^C dP_t^C + \xi_t^f dB_t^{r_f} + (1-\beta_t) \psi_t^{r_r} dB_t^{r_r} - \psi_t^c dB_t^{r_c}\\
=
& \Big( (1 -\beta_t \one_{ \{ \xi_t<0 \}}) r_D  \xi_t S_t + r_D \xi_t^I P_t^I + r_D \xi_t^C P_t^C  + r_f \xi_t^f B_t^{r_f} + (1-\beta_t) r_r \psi_t^{r_r} B_t^{r_r} - r_c\psi_t^c B_t^{r_c} \Big) dt\\
&+ (1-\beta_t \one_{ \{ \xi_t<0 \}}) \sigma \xi_t S_t dW_t^\BQ - \xi_t^IP_{t-}^I d\varpi_t^{I,\BQ} - \xi_t^C P_{t-}^C d\varpi_t^{C,\BQ}.\\
\end{split}
\end{equation*}

Then applying Assumption \ref{ass:repostock}, the above equation is rewrote as
\begin{equation*}
\begin{split}
dV_t 
=
& \Big( (1 -\beta_t \one_{ \{ \xi_t<0 \}}) r_D  \xi_t S_t + r_D \xi_t^I P_t^I + r_D \xi_t^C P_t^C  + r_f \xi_t^f B_t^{r_f} - (1-\beta_t) r_r \xi_t S_t - r_c\psi_t^c B_t^{r_c} \Big) dt\\
& + (1-\beta_t \one_{ \{ \xi_t<0 \}})  \sigma \xi_t S_t dW_t^\BQ - \xi_t^IP_{t-}^I d\varpi_t^{I,\BQ} - \xi_t^C P_{t-}^C d\varpi_t^{C,\BQ}\\
= 
&\Big( (r_D - r_D \beta_t \one_{\{ \xi_t < 0 \}} -r_r +r_r \beta_t) \xi_t S_t  + r_D \xi_t^I P_t^I + r_D \xi_t^C P_t^C  + r_f \xi_t^f B_t^{r_f} - r_c\psi_t^c B_t^{r_c} \Big) dt \\
&+ (1-\beta_t \one_{ \{ \xi_t<0 \}}) \sigma \xi_t S_t dW_t^\BQ - \xi_t^IP_{t-}^I d\varpi_t^{I,\BQ} - \xi_t^C P_{t-}^C d\varpi_t^{C,\BQ}.\\
\end{split}
\end{equation*}

Similarly, for the wealth process itself, we have that
$$
\xi_t^f B_t^{r_f} = V_t - \beta_t (1-\one_{\{ \xi_t < 0 \}})\xi_t S_t - \xi_t^I P_t^I - \xi_t^C P_t^C -C_t,\\
$$
by Assumption \ref{ass:repostock} and Equation \eqref{eq:collateral}.

Plugging the above equation into the dynamics of $V_t$, we get
\begin{equation*}
\begin{split}
dV_t = 
&\Big( (r_D - r_D \beta_t \one_{\{ \xi_t < 0 \}} -r_r +r_r \beta_t - r_f \beta_t + r_f \beta_t \one_{\{ \xi_t < 0 \}}) \xi_t S_t 
\\
& \quad + (r_D - r_f) \xi_t^I P_t^I + (r_D - r_r) \xi_t^C P_t^C   + r_f V_t + ( r_c - r_f) C_t \Big) dt 
\\
&\quad + (1-\beta_t \one_{ \{ \xi_t<0 \}}) \sigma \xi_t S_t dW_t^\BQ - \xi_t^IP_{t-}^I d\varpi_t^{I,\BQ} - \xi_t^C P_{t-}^C d\varpi_t^{C,\BQ}.
\end{split}
\end{equation*}
In order to simplify the above dynamics, we define three symbols as following.
\begin{equation}\label{eq:symbolz}
\begin{split}
Z_t &=  (1-\beta_t \one_{ \{ \xi_t<0 \}}) \sigma \xi_t S_t,\\
Z_t^I &= -\xi_t^IP_{t-}^I,\\
Z_t^C &= -\xi_t^C P_{t-}^C,
\end{split}
\end{equation} 

With these three symbols, second and third term of the drift equation in the dynamic of $V_t$ are represented as
\begin{equation}\label{zzizc1}
\begin{split}
(r_D - r_f) \xi_t^I P_t^I  &= - (r_D - r_f) Z_t^I,\\
(r_D - r_f) \xi_t^C P_t^C  &= -(r_D - r_f)Z_t^C.
\end{split}
\end{equation}

For the first term of the drift equation in the dynamic of $V_t$, we have
\begin{equation}\label{zzizc2}
\begin{split}
&(r_D - r_D \beta_t \one_{\{ \xi_t < 0 \}} -r_r +r_r \beta_t - r_f \beta_t + r_f \beta_t \one_{\{ \xi_t < 0 \}}) \xi_t S_t\\
=
&\frac{r_D}{\sigma}Z_t + (r_r \beta_t - r_f \beta_t + r_f \beta_t \one_{\{ \xi_t<0 \}} -r_r -r_f + r_f)\xi_t S_t\\
=
&\frac{r_D}{\sigma}Z_t - \frac{r_f}{\sigma} Z_t + (r_r \beta_t - r_f \beta_t -r_r + r_f)\xi_t S_t\\
=
&\frac{r_D}{\sigma}Z_t - \frac{r_f}{\sigma} Z_t + (r_f - r_r)(1 - \beta_t) \xi_t S_t.\\
\end{split}
\end{equation}

Next, we want to rewrite $(1-\beta_t) \xi_t S_t$ in a representation by $Z_t$.
By some comparison, we conclude that 
$$
(1-\beta_t)\xi_t S_t = \frac{1}{\sigma} (1-\one_{\{Z_t>0 , \beta_t =1\}}) Z_t.
$$
Based on this equation, we the last term in Equation \eqref{zzizc2} becomes
\begin{equation}\label{zzizc3}
(r_f - r_r) (1-\beta_t)\xi_t S_t = \frac{r_f - r_r}{\sigma} ( 1- \one_{\{ Z_t>0, \beta_t =1 \}}) Z_t .
\end{equation}

Plugging Equations \eqref{eq:symbolz}, \eqref{zzizc1}, \eqref{zzizc2}, \eqref{zzizc3} into the dynamics of $V_t$, we get that
\begin{equation*}
\begin{split}
dV_t 
= 
&\Big( \frac{r_D - r_f + (r_f - r_r) (1- \one_{\{ Z_t > 0, \beta_t = 1 \}})}{\sigma} Z_t - (r_D - r_f) Z_t^I - (r_D - r_f) Z_t^C + r_f V_t +( r_c - r_f)C_t \Big) dt \\
&\quad + Z_t dW_t^\BQ + Z_t^I d\varpi_t^{I,\BQ} + Z_t^C d\varpi_t^{C,\BQ}\\
= 
&\Big( r_f (V_t -\frac{\one_{\{ Z_t>0, \beta_t = 1 \}}}{\sigma}Z_t + Z_t^I + Z_t^C -C_t) - r_D Z^I - r_D Z^C + r_c C_t + (r_D - r_r(1-\one_{\{ Z_t>0, \beta_t = 1 \}}))\frac{Z_t}{\sigma} \Big)dt\\
&\quad + Z_t dW_t^\BQ + Z_t^I d\varpi_t^{I,\BQ} + Z_t^C d\varpi_t^{C,\BQ}\\
= 
&\Big( r_f^+ (V_t -\frac{ \one_{\{ Z_t>0, \beta_t = 1 \}}}{\sigma}Z_t + Z_t^I + Z_t^C -C_t)^+ - r_f^- (V_t -\frac{\one_{\{ Z_t>0, \beta_t = 1 \}}}{\sigma}Z_t + Z_t^I + Z_t^C -C_t)^-\\
&\quad+ \frac{r_D - r_r^-(1-\one_{\{ Z_t>0, \beta_t = 1 \}})}{\sigma} (Z_t)^+ - \frac{r_D - r_r^+(1-\one_{\{ Z_t>0, \beta_t = 1 \}})}{\sigma} (Z_t)^- \\
&\quad  + r_c^+ (\alpha \hat{V}_t)^+ - r_c^- (\alpha \hat{V}_t)^- - r_D Z^I - r_D Z^C \Big)dt \\
&\quad + Z_t dW_t^\BQ + Z_t^I d\varpi_t^{I,\BQ} + Z_t^C d\varpi_t^{C,\BQ}.\\
\end{split}
\end{equation*}

\subsection{Construction BSDEs of the Wealth Process}
In the dynamic of the wealth process $V_t$, the drift term has a very complicate structure. In order to simplify the equation, we define a generator function $f(t,v,z,z^I,z^C;\beta,\hat{V})$ as followings
\begin{equation*}
\begin{split}
f^+(t,v,z,z^I,z^C;\beta,\hat{V}) 
&= -\Big( r_f^+(v-\frac{\one_{\{ z>0, \beta=1 \}}}{\sigma} z +z^I + z^C - \alpha \hat{V})^+ -  r_f^-(v-\frac{\one_{\{ z>0, \beta=1 \}}}{\sigma} z +z^I + z^C - \alpha \hat{V})^-\\
& \quad + \frac{r_D - r_r^-(1-\one_{\{ z>0, \beta=1 \}})}{\sigma} z^+ - \frac{r_D - r_r^+(1-\one_{\{ z>0, \beta=1 \}})}{\sigma} z^- \\
&\quad + r_c^+(\alpha \hat{V})^+ - r_c^-(\alpha \hat{V}_t)^- - r_D z^I - r_D z^C\Big),\\
f^-(t,v,z,z^I,z^C;\beta,\hat{V}) & = -f^+ (t, -v, -z, -z^I,-z^C;\beta, -\hat{V}).
\end{split}
\end{equation*}

Moreover, we need the following assumptions.
\begin{ass}\label{assumptionf}
	We assume that
	\begin{enumerate}[(i)]
		\item $r_f^- < \frac{1}{5\sqrt{T^3}}$,
		\item $\frac{(r_f^- + r_D \vee |r_D - r_r^-|) \vee |r_D - r_r^+|}{\sigma \wedge 1} < \frac{1}{5T}$,
		\item $r_f^- - r_D < \frac{\sqrt{\lambda^I} \wedge \sqrt{\lambda^C}}{5T}$.
	\end{enumerate}	
\end{ass}

Now we construct two BSDEs with generator functions $f^\pm : \Omega\times[0,T] \times \RR^5 \times \{0,1\} \rightarrow \RR, \newline(\omega, t,v,z,z^I,z^C;\beta, \hat{V})\mapsto f^\pm(t,v,z,z^I,z^C;\beta, \hat{V})$ as
\begin{equation}\label{bsde1}
\left\{\begin{aligned}
-dV_t^+ &= f^+ (t, V^+_t, Z^+_t, Z^{I,+}_t, Z^{C,+}_t; \beta, \hat{V})dt -Z_t^+ dW_t^\BQ -Z_t^{I,+}d\varpi_t^{I,\BQ} - Z_t^{C,+} d\varpi_t^{C,\BQ},\\
V_\tau^+ &= \theta_I(\hat{V}_\tau) \one_{ \{ \tau^I < \tau^C \wedge T \} }  + \theta_C(\hat{V}_\tau) \one_{ \{\tau^C < \tau^I \wedge T\} } + \Theta \one_{\{\tau = T\}},
\end{aligned}
\right.
\end{equation}
and
\begin{equation}\label{bsde2}
\left\{\begin{aligned}
-dV_t^- &= f^- (t, V^-_t, Z^-_t, Z^{I,-}_t, Z^{C,-}_t; \beta, \hat{V})dt -Z_t^- dW_t^\BQ -Z_t^{I,-}d\varpi_t^{I,\BQ} - Z_t^{C,-} d\varpi_t^{C,\BQ},\\
V_\tau^- &= \theta_I(\hat{V}_\tau) \one_{ \{ \tau^I < \tau^C \wedge T \} }  + \theta_C(\hat{V}_\tau) \one_{ \{\tau^C < \tau^I \wedge T\} } + \Theta \one_{\{\tau = T\}},
\end{aligned}
\right.
\end{equation}
where $\hat{V}$ is a third party valuation  of $V$ with $\BE\big[ \int_0^T \hat{V}_s^2 ds \big] < \infty$ and $\Theta$\index{$\Theta$} is the terminal value at time $T$ without defaults. 

The process $V^+$ describes the wealth process to replicate the selling position of a claim with zero initial capital and a terminal payoff $\Theta$. Then, the process $V_t^-$ describes the wealth process to replicate the holding position of the claim. But in a financial crisis status, we can only super-hedge this claim, because of the frozen short selling trades of stock.

\subsection{Existence and Uniqueness of Solutions for the Valuation BSDEs}

In this section, we want to prove that there exists an unique solution of the valuation BSDE \eqref{bsde1} and BSDE \eqref{bsde2}. In order to prove it, we need a martingale decomposition theorem including non-independent increment process $\beta$. A literal statement of this theorem is listed here, but the mathematical version of this theorem is in Appendix \ref{chap:poisson}.

\begin{thm}\label{martdecomlit}
	Let $M$ be a $(\calF_t)_{t\geq 0}$ martingale with $\sup_{t \leq T}\BE[M_t^2] < \infty$. Then, there exists a unique decomposition of $M$ as 
	$$
	M_t = \int_0^t X_s dW_s + \int_0^t X_s^I d\varpi_s^I + \int_0^t X_s^C d\varpi_s^C + \int_0^t X_s^\beta d\tilde{J}_s + Y_t,
	$$
	where $X, X^I, X^C, X^\beta$ are $\mathscr{B}([0,t])\times \calF_t$ predictable processes with the square integrity, and $Y$ is orthogonal with all other stochastic integrals.
\end{thm}

We state the existence and uniqueness of solutions as the following theorem.

\begin{thm}\label{existbsdev}
	Given a filtered probability space $(\Omega, (\calF_t)_{t \geq 0}, \calF, \BQ)$, the BSDE \eqref{bsde1} admits an unique solution  $(V^+, Z^+, Z^{I,+}, Z^{C,+}) \in \BS^2 \times \mathscr{M}$. The solution satisfies the following equation	 
	\begin{equation*}
	\begin{split}
	V_t^+ &= V_\tau^+ + \int_t^\tau f^+(s,V_s^+, Z_s^+, Z_s^{I,+}, Z_t^{C,+}; \beta, \hat{V})ds  - \int_t^\tau Z^+_s dW^\BQ_s - \int_t^\tau Z_s^{I,+} d \varpi_t^{I,\BQ} - \int_t^\tau Z_s^{C,+} d\varpi_t^{C,\BQ}	.
	\end{split}
	\end{equation*}
\end{thm}

To prove Theorem \ref{existbsdev}, we need to prove the Lipschitz continuity of the generator function $f^+$ at first.

\begin{lem}\label{lipschitzf}
	For any given $0 < t < T, \beta, \hat{V}$, the generator functions $f^\pm$ are Lipschitz continuous in $v, z, z^I, z^C$ .
\end{lem}
\proof{Proof}
Given $\epsilon > 0$, $t \geq 0 $ and $\omega \in \Omega$, for  $ v_1, v_2, z_1, z_2, z^I_1, z^I_2, z^C_1, z^C_2$ with $|v_1 - v_2| < \epsilon, |z_1 - z_2| < \epsilon$, $|z^I_1 -z^I_2 | < \epsilon$ and $|z^C_1 - z^C_2 |< \epsilon$, we can prove this lemma by dividing it into three case.
\begin{enumerate}
	\item When $z_1, z_2$ are nonnegative, we have 
	
	\begin{equation*}
	\begin{split}
	&|f^+(t,v_1, z_1, z^I_1, z^C_1; \beta, \hat{V}) - f^+(t, v_2, z_2, z^I_2, z^C_2; \beta, \hat{V})| 		\\
	\leq 
	&~(r_f^+ \vee r_f^-)\Big| (v_2 - \frac{\one_{\{z_2>0, \beta = 1\}}}{\sigma}z_2 + z_2^I + z_2^C -\alpha \hat{V}_t) - (v_1 - \frac{\one_{\{z_1>0, \beta = 1\}}}{\sigma}z_1 + z_1^I + z_1^C -\alpha \hat{V}_t)\Big| \\
	& + \frac{r_D}{\sigma} |z_1^+ - z_2^+| + r_D |z_1^I - z_2^I| + r_D |z_1^C- z_2^C|\\
	\leq 
	&~ r_f^-\Big| (v_2 - \frac{\one_{\{z_2>0, \beta = 1\}}}{\sigma}z_2 + z_2^I + z_2^C -\alpha \hat{V}_t) - (v_1 - \frac{\one_{\{z_1>0, \beta = 1\}}}{\sigma}z_1 + z_1^I + z_1^C -\alpha \hat{V}_t)\Big| \\
	& + \frac{r_D}{\sigma} |z_1 -z_2| + r_D |z_1^I - z_2^I| + r_D |z_1^C- z_2^C|\\
	\leq 
	& ~r_f^- |v_1 - v_2| +  \frac{r_f^- + r_D}{\sigma} |z_1-z_2| +\left( r_f^- + r_D\right) |z^I_1 - z^I_2| +\left( r_f^- + r_D\right) |z^C_1 - z^C_2|\\
	\leq 
	&~ A_1 \left(|v_1 - v_2| + |z_1 - z_2| + |z_1^I - z_2^I| + |z_1^C- z_2^C|\right),
	\\
	\end{split}
	\end{equation*}
	where $A_1 = \frac{r_f^- + r_D}{\sigma \wedge 1}$.
	
	\item When $z_1, z_2$ are negative, we have 
	
	\begin{equation*}
	\begin{split}
	&|f^+(t,v_1, z_1, z^I_1, z^C_1; \beta, \hat{V}) - f^+(t, v_2, z_2, z^I_2, z^C_2; \beta, \hat{V})| 		\\
	\leq 
	&~(r_f^+ \vee r_f^-)\Big| (v_2 + z_2^I + z_2^C -\alpha \hat{V}_t ) - (v_1 + z_1^I + z_1^C -\alpha \hat{V}_t)\Big| \\
	& +  \frac{| r_D - r_r^+|}{\sigma } |z_1^- - z_2^-| + r_D |z_1^I - z_2^I| + r_D |z_1^C- z_2^C|
	\\
	\leq 
	& ~r_f^-| v_1 - v_2| + \frac{|r_D - r_r^+|}{\sigma}|z_1 - z_2| + (r_f^- + r_D) |z_1^I - z^I_2| + (r_f^- + r_D) |z_1^C - z^C_2|\\
	\leq &~ A_2 \left(|v_1 - v_2| + |z_1 - z_2| + |z_1^I - z_2^I| + |z_1^C- z_2^C|\right),
	\\
	\end{split}
	\end{equation*}
	where $A_2 = (r_f^-+ r_D) \vee \frac{ |r_D - r_r^+| }{\sigma\wedge 1}$.
	
	\item Without loss of generality, we assume that $z_1 > 0$ and $ z_2 < 0$, we have $|z_1 + z_2| \leq |z_1 + |z_2|| = |z_1 - z_2| < \epsilon$ and $|z_1| \leq |z_1 + |z_2|| = |z_1 - z_2| < \epsilon$.
	
	\begin{equation*}
	\begin{split}
	&|f^+(t,v_1, z_1, z^I_1, z^C_1; \beta, \hat{V}) - f^+(t, v_2, z_2, z^I_2, z^C_2; \beta, \hat{V})| 		\\
	\leq 
	&~(r_f^+ \vee r_f^-)\Big| (v_2 + z_2^I + z_2^C -\alpha \hat{V}_t ) - (v_1 - \frac{\one_{\{z_1>0, \beta = 1\}}}{\sigma}z_1 + z_1^I + z_1^C -\alpha \hat{V}_t )\Big| \\
	& + \Big| \frac{r_D - r_r^-(1-\one_{\{ z_1>0, \beta=1 \}})}{\sigma} z^+_1 + \frac{r_D - r_r^+}{\sigma} z^-_2 \Big| + r_D |z_1^I - z_2^I| + r_D |z_1^C- z_2^C|
	\\
	\leq 
	&~r_f^-\Big| (v_2 + z_2^I + z_2^C -\alpha \hat{V}_t ) - (v_1 - \frac{\one_{\{z_1>0, \beta = 1\}}}{\sigma}z_1 + z_1^I + z_1^C -\alpha \hat{V}_t )\Big| \\
	& + \frac{r_D \vee |r_D - r_r^+|}{\sigma}| z_1 + |z_2| | + r_D |z_1^I - z_2^I| + r_D |z_1^C- z_2^C|
	\\
	\leq 
	&~ r_f^- |v_1 - v_2| + \frac{r_f^-}{\sigma} |z_1|+ \frac{r_D \vee |r_D - r_r^-|}{\sigma} |z_1 + |z_2|| + (r_f^- + r_D) |z^I_1 - z^I_2| + (r_f^- + r_D)|z^C_1 - z^C_2|
	\\
	\end{split}
	\end{equation*}
	\begin{equation*}
	\begin{split}
	\leq 
	& ~r_f^- |v_1 - v_2| + \frac{r_f^-}{\sigma} |z_1 - z_2|+ \frac{r_D \vee |r_D - r_r^-|}{\sigma} |z_1 - z_2| + (r_f^- + r_D) |z^I_1 - z^I_2| + (r_f^- + r_D)|z^C_1 - z^C_2|
	\\
	\leq 
	&~ r_f^- |v_1 - v_2| +  \frac{r_f^- + (r_D \vee |r_D - r_r^-|)}{\sigma} |z_1 - z_2| + (r_f^- + r_D) |z^I_1 - z^I_2| + (r_f^- + r_D)|z^C_1 - z^C_2|
	\\
	\leq 
	&~ A_3 \left(|v_1 - v_2| + |z_1 - z_2| + |z_1^I - z_2^I| + |z_1^C- z_2^C|\right),
	\\
	\end{split}
	\end{equation*}
	where $A_3 = \frac{r_f^- + (r_D \vee |r_D - r_r^-|)}{\sigma\wedge 1} \vee (r_f^- + r_D)$.  
	
\end{enumerate}
Overall, the function  $f^+$ satisfies the Lipschitz condition in $v, z, z^I, z^C$ such that
$$
|f^+(t,v_1, z_1, z^I_1, z^C_1; \beta, \hat{V}) - f^+(t, v_2, z_2, z^I_2, z^C_2; \beta, \hat{V})|  \leq K \left(|v_1 - v_2| + |z_1 - z_2| + |z_1^I - z_2^I| + |z_1^C- z_2^C|\right),
$$
where $K = A_1 \vee A_2 \vee A_3 = \frac{(r_f^- + r_D \vee |r_D - r_r^-|) \vee |r_D - r_r^+| }{\sigma\wedge 1}$ independently for $\beta$ and $\hat{V}$.
\endproof

We already proved the Lipschitz condition of the generator function $f^+$, we will prove the Theorem \ref{existbsdev} here by Lemma \ref{lipschitzf} and Theorem \ref{martdecomlit}.

\proof{Proof of Theorem \ref{existbsdev}}
~\\
To prove this theorem, BSDE  \eqref{bsde1} has to satisfies the Lipschitz Condition, Terminal Condition and Integrability Condition of Assumption \ref{existuniqueass} in Appendix \ref{chap:poisson}.
\begin{enumerate}
	\item Lipschitz Condition
	
	By the Lemme \ref{lipschitzf}, we have the generator function $f^+$ has the inequality
	\begin{equation*}
	\begin{split}
	&|f^+(t,v_1, z_1, z^I_1, z^C_1; \beta, \hat{V}) - f^+(t, v_2, z_2, z^I_2, z^C_2; \beta, \hat{V})|\\
	\leq
	&  r_f^-|v_1 - v_2| + \frac{(r_f^- + r_D \vee |r_D - r_r^-|) \vee |r_D - r_r^+|}{\sigma}|z_1 - z_2| + (r_f^- + r_D)|z_1^I - z_2^I| + (r_f^- + r_D)|z_1^C- z_2^C| .
	\end{split}		
	\end{equation*} 
	By Assumptions \ref{assumptionf}, the above inequality becomes
	\begin{equation*}
	\begin{split}
	&|f^+(t,v_1, z_1, z^I_1, z^C_1; \beta, \hat{V}) - f^+(t, v_2, z_2, z^I_2, z^C_2; \beta, \hat{V})|
	\\
	<
	& ~ \frac{1}{5\sqrt{T^3}}|v_1 - v_2| + \frac{\sqrt{\lambda^I}}{5T}|z_1^I - z_2^I| + \frac{\sqrt{\lambda^C}}{5T}|z_1^C- z_2^C| + \frac{1}{5T}|z_1 - z_2|
	\\
	=
	& ~ \frac{1}{5T}\left(\frac{1}{\sqrt{T}}|v_1 - v_2| + |z_1 - z_2| + \sqrt{\lambda^I}|z^I_1 - z^I_2| + \sqrt{\lambda^C}|z^C_1 - z^C_2|\right).
	\end{split}		
	\end{equation*} 
	So our generator function $f^+$ satisfies the Lipschitz condition in Assumption \ref{existuniqueass}.
	
	\item Terminal Condition
	
	By the definition of $V_\tau$, we have $V_\tau \in L^2(\Omega, \calF_\tau, \BQ)$, so the closeout valuation satisfies the terminal condition in Assumption \ref{existuniqueass}.
	
	\item Integrability Condition
	
	By the definition of $f^+$, we have 
	\begin{equation*}
	f^+(t,0,0,0,0;\beta,\hat{V}) = -\Big( r_f^+(- \alpha \hat{V}_t)^+ - r_f^-(-\alpha \hat{V}_t)^- + r_c^+(\alpha \hat{V}_t)^+ - r_c^-(\alpha \hat{V}_t)^- \Big).
	\end{equation*}
	For $T < \infty$, 
	$$
	\BE \Big[ \int_0^T |f^+(s,0,0,0,0;\beta,\hat{V})|^2 ds\Big] < \infty,
	$$ 
	so $f^+(t,0,0,0,0;\beta,\hat{V})\in \HH^2$ satisfies the integrability condition in Assumption \ref{existuniqueass}.
\end{enumerate}

By Theorem \ref{martdecomlit} and Theorem \ref{thmbsde} in Appendix \ref{chap:poisson}, this valuation BSDE \eqref{bsde1} with the generator $f^+$ admits a unique solution $(V, Z, Z^I, Z^C,Z^\beta, Y) \in \BS^2 \times \mathscr{M}$. 
\begin{equation*}
\begin{split}
V_t^+ &= V_\tau^+ + \int_t^\tau f(\omega, s, V_{s-}, Z_s, Z^I_s, Z^C_s, Z^\beta_s) ds + \int_t^\tau Z_s dW_s^\BQ + \int_t^\tau Z^I_s d \varpi_t^{I,\BQ} + \int_t^\tau Z^C_s d\varpi_t^{C,\BQ} \\
&\qquad + \int_t^\tau Z^\beta_s d\tilde{J}_s + Y_\tau - Y_t.
\end{split}
\end{equation*}

Based on the specific form of the valuation BSDE with the generator $f^+$, the solution does not depend on the stochastic integral with respect to $\tilde{J}$ and the orthogonal term $Y$. So the solution $(V,Z,Z^I,Z^C)$ to BSDE \eqref{bsde1} is given by
$$
V_t^+ = V_\tau^+ + \int_t^\tau f^+(s, V_s^+, Z_s^+, Z^{I,+}_s, Z^{C,+}_s; \beta, \hat{V}) ds - \int_t^\tau Z^+_s dW_s^\BQ - \int_t^\tau Z^{I,+}_s d \varpi_t^{I,\BQ} - \int_t^\tau Z^{C,+}_s d\varpi_t^{C,\BQ}.
$$

\endproof

\begin{thm}\label{existbsdev2}
	Given $(\Omega, (\calF_t)_{t \geq 0}, \calF, \BP)$, BSDE \eqref{bsde2} admits a unique solution \newline $(V^-, Z^-, Z^{I,-}, Z^{C,-}) \in \BS^2 \times \mathscr{M}$. The solutions satisfies the following equation
	{\small	 
		\begin{equation*}
		V_t^- = V_\tau^- + \int_t^\tau f^-(s,V_s^-, Z_s^-, Z_s^{I,-}, Z_t^{C,-}; \beta, \hat{V})ds  - \int_t^\tau Z^-_s dW^\BQ_s - \int_t^\tau Z_s^{I,-} d \varpi_t^{I,\BQ} - \int_t^\tau Z_s^{C,-} d\varpi_t^{C,\BQ}.
		\end{equation*}}
\end{thm}

\proof{Proof}
The proof is similar to the proof of Theorem \ref{existbsdev}.
\endproof


\section{Total Valuation Adjustment (XVA)}\label{sec:xva}
In this section, we develop a BSDEs framework to price the XVA of one European option and prove the existence and uniqueness of its solution. 
The total valuation adjustment (XVA) is an adjustment made to the fair value of a derivative contract to take into account funding and credit risk. We  want to compute the total valuation adjustment, which is added to the Black-Scholes price of a European option, considering an investor's and its counterparty's defaults, funding liquidity and asymmetric interest rates.

\subsection{Construction of XVA BSDEs }
Because the valuation adjustments are asymmetric for the buy-side and sell-side, we define both seller's and buyer's XVA here.
\begin{dfn}\label{def:xva}
	The seller's XVA is a value adjustment between the seller's valuation of the claim, compared with a third party valuation $\hat{V}$, which is a stochastic process defined as
	$$
	\text{XVA}_t^+ := V_t^+ - \hat{V}_t.
	$$
	Similarly, the buyers' XVA is the total valuation adjustment between the buyer's valuation and a third party valuation $\hat{V}$, which is a stochastic process defined as
	$$
	\text{XVA}_t^- : = V_t^- - \hat{V}_t.
	$$
\end{dfn}
$\text{XVA}^+$ is a valuation adjustment by the trader to hedge a long position in the option, and $\text{XVA}^-$ is a valuation adjustment by the trader to hedge a short position in the option.

In this paper, $\hat{V}$ is a solution of the Black-Scholes model
\begin{equation}
\left\{\begin{aligned}
-d\hat{V}_t &= -r_D \hat{V}_t dt - \hat{Z}_t dW_t^\BQ,\index{$\hat{V}$}\\
\hat{V}_T &= \Theta,	
\end{aligned}\right.
\end{equation}
where $\Theta$ is the terminal value of the claim from the point of view the third party.

In Section \ref{sec:pricing}, we constructed a BSDEs framework to evaluate the price of one claim, considering the defaults from the investor and its counterparty, the switching between different financial regimes and asymmetric interest rates. By Definition \ref{def:xva} and BSDEs \eqref{bsde1} \eqref{bsde2}, we have the BSDEs for the $\text{XVA}^\pm$ as
\begin{equation}\label{xvabsde}
\left\{\begin{aligned}
-d\text{XVA}_t^\pm &= \tilde{f}^\pm (t, \text{XVA}^\pm_t, \tilde{Z}^\pm_t, \tilde{Z}^{I,\pm}_t, \tilde{Z}^{C,\pm}_t; \beta, \hat{V}, 
\hat{Z})dt -\tilde{Z}_t^\pm dW_t^\BQ -\tilde{Z}_t^{I,\pm}d\varpi_t^{I,\BQ} - \tilde{Z}_t^{C,\pm} d\varpi_t^{C,\BQ},\\
\text{XVA}_\tau^\pm &= \tilde{\theta}_I(\hat{V}_\tau) \one_{ \{ \tau^I < \tau^C \wedge T \} }  + \tilde{\theta}_C(\hat{V}_\tau) \one_{ \{\tau^C < \tau^I \wedge T\} },
\end{aligned}\right.
\end{equation}
where 
\begin{equation}
\begin{split}
\tilde{Z}^\pm_t &:= Z_t^\pm - \hat{Z}_t,\\
\tilde{Z}_t^{I, \pm} &:= Z_t^{I, \pm},\\
\tilde{Z}_t^{C, \pm} &:= Z_t^{C, \pm}, \\
\tilde{\theta}_I(\hat{v}) &:= -L_I((1-\alpha)\hat{v})^+,\\
\tilde{\theta}_C(\hat{v}) &:= L_C((1-\alpha)\hat{v})^-, 
\end{split}
\end{equation}
and the generators are 
\begin{equation}
\begin{split}
&\tilde{f}^+(t,xva,\tilde{z},\tilde{z}^I,\tilde{z}^C;\beta,\hat{V},\hat{Z}) \\
= &-\Big( r_f^+(xva-\frac{\one_{\{ \tilde{z}+\hat{Z}>0, \beta=1 \}}}{\sigma} (\tilde{z}+\hat{Z}) +\tilde{z}^I + \tilde{z}^C +(1 - \alpha) \hat{V})^+  + \frac{r_D - r_r^-(1-\one_{\{ \tilde{z}+\hat{Z}>0, \beta=1 \}})}{\sigma} (\tilde{z}+\hat{Z})^+ \\
&\qquad -r_f^-(xva-\frac{\one_{\{ \tilde{z}+\hat{Z}>0, \beta=1 \}}}{\sigma} (\tilde{z}+\hat{Z}) +\tilde{z}^I + \tilde{z}^C +(1- \alpha) \hat{V})^- - \frac{r_D - r_r^+(1-\one_{\{ \tilde{z}+\hat{Z}>0, \beta=1 \}})}{\sigma} (\tilde{z}+\hat{Z})^- \\
&\qquad + r_c^+(\alpha \hat{V}_t)^+ - r_c^-(\alpha \hat{V}_t)^- - r_D \tilde{z}^I - r_D \tilde{z}^C\Big) +r_D\hat{V}_t,\\
&\tilde{f}^-(t,xva,\tilde{z},\tilde{z}^I,\tilde{z}^C;\beta,\hat{V},\hat{Z})  := -\tilde{f}^+ (t, -xva, -\tilde{z}, -\tilde{z}^I,-\tilde{z}^C;\beta, -\hat{V},-\hat{Z}).
\end{split}
\end{equation}
Based on the definition of the generator function of BSDEs \eqref{bsde1} and \eqref{bsde2}, we have
\begin{equation}\label{ftildef}
\tilde{f}^\pm(t,xva,\tilde{z},\tilde{z}^I,\tilde{z}^C;\beta,\hat{V},\hat{Z}) = f^\pm(t,xva,\tilde{z}+\hat{Z},\tilde{z}^I,\tilde{z}^C;\beta, \hat{V})\pm r_D\hat{V}.
\end{equation}

\begin{thm}\label{existbsdexva}
	Given a filtered probability space $(\Omega, (\calF_t)_{t \geq 0}, \calF, \BQ)$. The XVA BSDEs \eqref{xvabsde}
	admit a unique solution $(\text{XVA}^\pm, \tilde{Z}^\pm, \tilde{Z}^{I,\pm}, \tilde{Z}^{C,\pm})$. The solution satisfies the following equation 	 
	{\small
		\begin{equation*}
		\text{XVA}_t^\pm = \text{XVA}_\tau^\pm + \int_t^\tau \tilde{f}^\pm(s,\text{XVA}_s^\pm, \tilde{Z}_s^\pm, \tilde{Z}_s^{I,\pm}, \tilde{Z}_t^{C,\pm}; \beta, \hat{V}, \hat{Z})ds  - \int_t^\tau \tilde{Z}^+_s dW^\BQ_s - \int_t^\tau \tilde{Z}_s^{I,\pm} d \varpi_t^{I,\BQ} - \int_t^\tau \tilde{Z}_s^{C,\pm} d\varpi_t^{C,\BQ}.
		\end{equation*}}
\end{thm}

\proof{Proof}
The result is a direct consequence of Theorem \ref{existbsdev} and the Black-Scholes formula.
\endproof

\subsection{Reduced XVA BSDEs}

The XVA BSDEs \eqref{xvabsde} has a BSDEs witha jump terminal condition in the filtration $(\calF_t)_{t \geq 0}$. In this paper, we want to rewrite the XVA BSDEs \eqref{xvabsde} to a reduced XVA BSDEs with a smaller filtration $(\calF^{W,\beta})_{t \geq 0}$ and a continuous terminal condition. The reduced XVA BSDEs are
\begin{equation}\label{redxvabsde}
\left\{\begin{aligned}
- d\breve{U}_t^\pm \index{$ \breve{U}^\pm$} & = \breve{g}^\pm (t, \breve{U}_t^\pm, \breve{Z}_t^\pm; \beta, \hat{V}, \hat{Z}) dt - \breve{Z}_t^\pm dW^\BQ_t,\\
\breve{U}_T^\pm & = 0,	
\end{aligned}
\right.
\end{equation}

in the filtration $(\calF^{W, \beta}_t)_{t \geq 0}$ without default events with\index{$\breve{g}^\pm$}
\begin{equation}\label{brevegdef}
\begin{split}
\breve{g}^+ (t,\breve{u}, \breve{z}; \beta, \hat{V}, \hat{Z}) & := h_I^\BQ(\tilde{\theta}_I(\hat{V}_t) - \breve{u}) + h_C^\BQ (\tilde{\theta}_C(\hat{V}_t) - \breve{u}) + \tilde{f}^+ (t,\breve{u},\breve{z}, \tilde{\theta}_I(\hat{V}_t) - \breve{u}, \tilde{\theta}_C(\hat{V}_t) - \breve{u}; \beta, \hat{V}, \hat{Z}),\\ 
\breve{g}^-(t, \breve{u}, \breve{z}; \beta, \hat{V}, \hat{Z}) &: = - \breve{g}^+ (t, - \breve{u}, -\breve{z}; \beta, -\hat{V}, -\hat{Z}) .
\end{split}
\end{equation}

\begin{thm}\label{redbsde}
	The reduced XVA BSDE \eqref{redxvabsde} admits a unique solution $(\breve{U}^\pm, \breve{Z}^\pm)$. When $ ( \text{XVA}^\pm, \tilde{Z}^\pm, \tilde{Z}^{I,\pm}, \tilde{Z}^{C, \pm} ) $ is a unique solution of  BSDEs \eqref{xvabsde}, then $(\breve{U}^\pm, \breve{Z}^\pm)$ defined as
	$$
	\breve{U}_t^\pm := \text{XVA}^\pm_{t\wedge \tau-}, \quad
	\breve{Z}_t^\pm := \tilde{Z}_t^\pm \one_{\{ t < \tau \}},	
	$$
	are solutions to the reduced XVA BSDE \eqref{redxvabsde}.	
	When $(\breve{U}^\pm, \breve{Z}^\pm)$ are unique solutions to the reduced XVA BSDEs \eqref{redxvabsde}, then $(\text{XVA}^\pm, \tilde{Z}^\pm, \tilde{Z}^{I,\pm}, \tilde{Z}^{C, \pm})$, defined as
	\begin{equation*}
	\begin{split}
	&\text{XVA}_t^\pm := \breve{U}_t^\pm \one_{\{ t<\tau \}} + \Big(\tilde{\theta}_I(\hat{V}_{\tau_I}) \one_{ \{ \tau_I < \tau^C \wedge T \} }  + \tilde{\theta}_C(\hat{V}_{\tau_C}) \one_{ \{\tau_C < \tau^I \wedge T\} } \Big) \one_{\{ t\geq \tau \}},
	\\
	&\tilde{Z}_t^\pm  := \breve{Z}_t^\pm \one_{\{ t < \tau \}}, \quad	\tilde{Z}^{I,\pm}_t := \Big( \tilde{\theta}_I(\hat{V}_t) - \breve{U}_t^\pm \Big)\one_{\{ t \leq \tau \}}, \quad	\tilde{Z}^C_t := \Big(\tilde{\theta}_C(\hat{V}_t) - \breve{U}_t^\pm \Big)\one_{\{ t \leq \tau \}},
	\end{split}
	\end{equation*}
	are unique solutions of the XVA BSDEs \eqref{xvabsde}.
\end{thm}

\proof{Proof}
Similary to the proof of Theorem \ref{existbsdev}, we need the Lipschitz condition of the generator functions $\breve{g}^\pm$. By Equation \eqref{brevegdef}, it's obvious.
Based on the definition of $\breve{g}^\pm$, the integrability and terminal conditions are trivial. By Theorem \ref{thmbsde}, the reduced XVA BSDEs \eqref{redxvabsde} admit a unique solution.

The equivalence between the XVA BSDEs \eqref{xvabsde} and the reduced XVA BSDEs \eqref{redxvabsde} follows from Theorem 4.3 in \cite{crepey2015bsdes} and Theorem \ref{martdecom}. 

\endproof

\section{Empirical Application}\label{sec:pricing_simu}
In this section, we illustrate some simulation results of the alternating renewal process $\beta$ and the XVA valuation of a European call option. We want to estimate the parameters of the alternating renewal process by using historical data and compare the XVA with and without considering the different performances of the Repo account and stock account during different financial statuses.

\subsection{Estimations of Alternating Renewal Processes}

In this section, we want to estimate the parameters $\lambda_U$ and $\lambda_V$ of the alternating renewal process $\beta$, which are expected lengths of a normal financial regime and a financial crisis. So we need to select historical data from some financial stress index to estimate these parameters. As we reviewed in Section \ref{chap:intro}, there are several indicators of financial distress, such as VIX, CoVaR and the Ted spread. For Ted spread, \cite{boudt2017funding} confirm the existence of a two-regime Ted spreads from January 2006 to December 2011. Their estimation of the threshold for the regime switching is 0.48 basis points. Here we also use the historical data of the Ted spread to estimate the parameters $\lambda_U$ and $\lambda_V$. 

\begin{figure}[h]
	\centering
	\subfloat[Threshold 0.48.]{{\includegraphics[width=.45\textwidth]{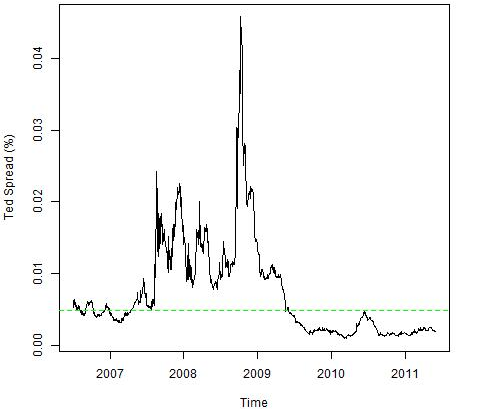} }}\label{fig:tedspread48}
	\qquad
	\subfloat[Thresholds 0.48 and 0.8.l 2]{{\includegraphics[width=.45\textwidth]{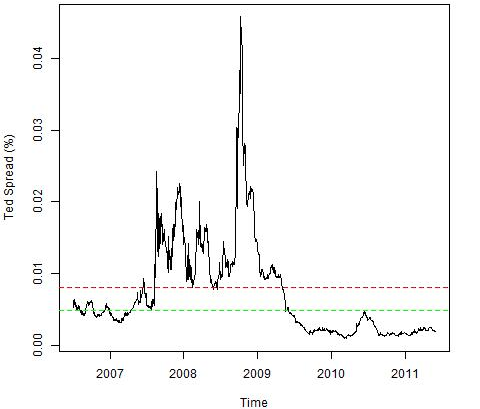} }}\label{fig:tedspread4880}
	\caption{Ted spread from Jan 2006 to Dec 2011.}%
\end{figure}

Setting the threshold at .48 basis points,  we claim that the financial market enters a crisis status when the Ted spread is larger than the 48 basis points. When the Ted spread is smaller or equal to the .48 basis points, we claim that the financial market is in a normal status. Assuming all financial status are independent, we use the sample mean to estimate the expectation of the lengths of the financial statuses. The results are given in Table \ref{table:lambda48}. In this result, the estimates are relatively small. In Figure \ref{fig:tedspread48}, we can see that the Ted spread crossed the threshold (green line) 10 times, which means both statuses appeared five times in the dataset. But this number contradicts to the real financial condition from 2006 to 2011. The cause of this contradiction is the noise information during a financial stress period. The Ted spread data represents those noise information, so there were more several oscillation at the beginning of the financial crisis. 

\begin{table}[!h]
	\centering
	\begin{tabular}{|c|c|c|}
		\hline
		~	& Normal Financial Statuses  & Financial Crisis Statuses \\
		\hline
		\hline
		Number of Statuses  &  5  & 5 \\
		\hline
		Average Length (days)  & 179 &	172	\\
		\hline
		Estimates for $\lambda_U$ and $\lambda_V$  & 0.49 &	0.47 \\
		\hline   
	\end{tabular}
	\caption{ Estimates of $\lambda_U$ and $\lambda_V$ when threshold is .48 basis points. }
	\label{table:lambda48}
\end{table}

In order to remove the effect of noise information, \cite{boudt2013funding} mention that the 0.8 basis points is also a meaningful threshold, which is the threshold for the central bank responds to a financial crisis. We combine these two thresholds 0.48 and 0.8 basis points to eliminate the effect of the Ted spread's noise movements. When the Ted spread up-crosses the .80 basis points (red line in Figure \ref{fig:tedspread4880}), we claim that the financial market enters a financial crisis. When the Ted spread down-crosses the .48 basis points (green line in Figure \ref{fig:tedspread4880}), we claim that the financial market enters a normal financial regime. 
Based on the setting of .48 and .80 basis points, we estimate the parameters $\lambda_U$ and $\lambda_V$ of the alternating renewal process $\beta$, given in Table \ref{table:lambda4880}.

\begin{table}[!h]
	\centering
	\begin{tabular}{|c|c|c|}
		\hline
		~	& Normal Financial Statuses & Financial Crisis Statuses \\
		\hline
		\hline
		Number of Statuses  &  2  & 2 \\
		\hline
		Average Length (days)  & 507 &	361	\\
		\hline
		Estimates of $\lambda_U$ and $\lambda_V$  & 1.39 &	0.99
		\\
		\hline   
	\end{tabular}
	\caption{ Estimations of $\lambda_U$ and $\lambda_V$ when threshold is .48 and .8 basis points. }
	\label{table:lambda4880}
\end{table}

Besides using the Ted spread, we can also use other financial stress indicators. It would be interesting to analyze these financial indicators statistically and find reasonable thresholds of different financial regimes in future.

\subsection{Simulation Results of XVAs}
In this section, we want to simulate the $\text{XVA}^+$
Assume that an initial stock price $S_0 = \$1$, the strike price $K = \$1$ and the terminal time of the option is three months ($T = 0.25$ year). We set the following benchmark coefficients: $r_r^+ = r_r^- = 0.05$, $r_c^+ = r_c^- = 0.01$,  $r_f^+ = 0.05, r_D = 0.01, \mu^I = 0.21, \mu^C = 0.16, \sigma = 0.3, h_I^\BQ = 0.2, h_C^\BQ = 0.15$, and $L_I = L_C = 0.5$, as in \cite{bichuch2018arbitrage}. Firstly, we want to analyze the effect of the financial regimes ($\beta$), the different collateralization levels $\alpha$ and the borrowing funding rates ($r_f^-$) on the XVA. Then we want to analyze the change of XVA, corresponding of different estimation of $\lambda_U$.

\begin{figure}[h]
	\centering
	\subfloat[$\beta_t = 0$.]{{\includegraphics[width=.45\textwidth]{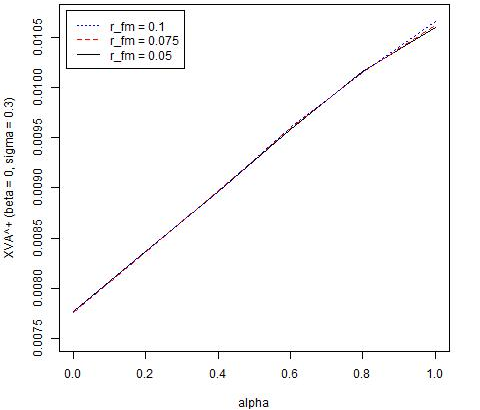} }}\label{fig:xvabeta0sigma2}
	\qquad
	\subfloat[$\beta_t = 1$.]{{\includegraphics[width=.45\textwidth]{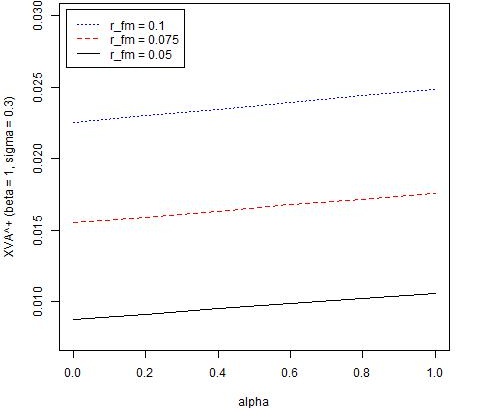} }}\label{fig:xvabeta1sigma2}
	\caption{$\text{XVA}^+$ when $\sigma = 0.3$}	
\end{figure}

In this paper, we use the deep learning-based numerical method to solve our XVA BSDE, the algorithm can be found in \cite{weinan2017deep}. We set the learning rate as 5e-6 and trainig steps as 3e6. During a normal financial regime ($\beta_t = 0$), we compute the XVA$^+$ for different collateralization levels $\alpha$ between $0$ and $1$ and different funding rates $r_f^- = 0.05, 0.075,$ and $0.1$, the result is plotted in Figure \ref{fig:xvabeta0sigma2}. The $\text{XVA}^+$ increases corresponding to the increasement in the collateralization level $\alpha$. The increments of the XVA$^+$ for the different funding rates $r_f^-$ increase slightly for collateralization levels. This result means that the hedger invest their money from selling stocks in the Repo market, which is consistent with the no-arbitrage assumption $r_r^+ > r_f^+$. When $\alpha$ increases, the investor needs to hold more shares in the collateral account $C_t$, which leads to borrowing more cash from the funding account. In this situation, the XVA also increases due to the higher funding cost incurred  the hedging of the investor's and its counterparty's default risks. However, in a financial crisis, the XVA$^+$ increase sharply corresponding with the increase of $r_f^-$. Because the Repo market freezes during a financial crisis, all borrowing of cash has to be ceased through the funding market. Therefore, the increasement in the funding rate $r_f^-$ effect the $XVA^+$ dramatically. The relationship between XVA$^+$ and the collateralization level $\alpha$ is simplified to a linear relationship. The results are plotted in Figure \ref{fig:xvabeta1sigma2}.
Compared XVA$^+$ for different financial regimes,the XVA$^+$ in a crisis is nearly double the size of the XVA$^+$ in a normal financial regime. Therefore, it is important to differentiate the different financial statuses when pricing an option.

\begin{figure}[!h]
	\centering
	\includegraphics[width= 0.4\linewidth]{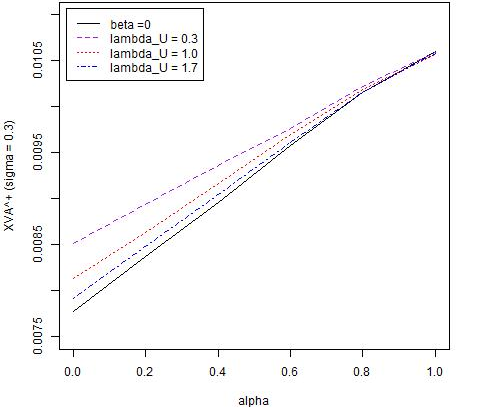}
	\caption{ $XVA^+$ for $\beta_t = 0$ and a dynamic beta process.}
	\label{fig:xvalambdaU}
\end{figure}

We also compare the XVA$^+$ with different estimations of $\lambda_U$ and collateralization levels $\alpha$, given in Figure \ref{fig:xvalambdaU}. Since $\lambda_U$ is the expectation length of the normal financial regimes, the larger $\lambda_U$ means a longer stable financial regime. This result shows that investor need larger XVA$^+$ to hedge the incoming financial crisis, especially when a financial crisis is close (smaller $\lambda_U$). However, when the collateral account is a fully collateralized account, the XVA$^+$ doesn't change corresponding to an incoming financial crisis. Since we assume the collateralization in collateral account is cash, we ignore the fire sales in this paper. In future, it's interest to include the market illiquidity problem into the collateral account. In that case, XVA maybe change corresponding with the change of the estimation of $\lambda_U$ when $\alpha = 1$.

\section{Conclusion}\label{chap:con}

In this paper, we discussed pricing of options emerging in the wake of the 2008 financial crisis. We provided the arbitrage free pricing of one claim, considering credit risk, asymmetric interest rates and the different performances of several financial accounts during different financial statuses. 

To model the frozen bilateral Repo market and short-selling of stocks during the financial crisis, we used an alternating renewal process to describe the switching between different financial regimes. With a replicating portfolio including risky bonds from the investor and its counterparty, we constructed a BSDE with respect to a martingale without independent increments property to price the claim and the corresponding XVA. We proved the existence and uniqueness of the solution to these BSDEs. In the empirical application, we estimated the length of different financial periods by Ted spread historical data. We also analyzed the sensitivity of the XVA to collateralization levels, expected length of a normal financial regime ($\lambda_U$) and the borrowing funding rates ($r_f^-$). In a simulation study, the XVA in the financial crisis increased 100\%, compared with the XVA in a calm financial market.

\appendix
	
	\section{Alternating Renewal Process $\beta$}\label{chap:poisson}
	
	By the definition of the alternating renewal process, the stochastic process $\beta$ is a right continuous Markov process with left limits ($c\grave{a}dl\grave{a}g$), switching between status $``0"$ and status $``1"$. At each alternating time $T_n$, the jump direction of the process $\beta$ depends on the status of $\beta_{T_n -}$. When $\beta_{T_n -} = 0$, then $\beta_{T_n} = 1$ as a result of an upward jump. When $\beta_{T_n-} = 1$, then $\beta_{T_n} = 0$ as a result of a downward jump. Therefore, the alternating renewal process $\beta$ does not have an independent increments property. Because $\beta$ has only countable many jumps almost surely, it has one $c\grave{a}gl\grave{a}d$ modification. We decompose $\beta$ as the sum of two processes $\beta^+$ and $\beta^-$:
	\begin{equation}\label{beta+beta-}
	\beta^+_t = \sum_{s\leq t} \one_{\{ \beta_{s-} < \beta_s\}}, 
	\qquad \beta^-_t = - \sum_{s\leq t} \one_{\{ \beta_{s-} > \beta_s\}}.
	\end{equation}
	
	The process $\beta^+$\index{$\beta^\pm$} is a jump counting process of the upward jumps and the process $\beta^-$ is a jump counting process of a downward jumps.
	
	Since $\beta^-$ is a non-increasing $c\grave{a}dl\grave{a}g$ process, it is a supermartingale. By the definition, the inter-arrival times of $\beta^-$ are exponential distributed random variables with parameter $\lambda$, where $\lambda = \frac{\lambda_U \lambda_V}{\lambda_U + \lambda_V}$.
	In the same way, we know that $\beta^+$ is a submartingale. Contrary to $\beta^-$, it does not have the independent increments property. It is a sum of one exponential process with parameter $\lambda_U$ and a Gamma process $ Gamma(n,\lambda)$, where $\lambda = \frac{\lambda_U \lambda_V}{\lambda_U + \lambda_V}$. 
	We get the probability distribution function of $\beta^+_t$ as follows:
	
	\begin{equation}\label{eq:probbetap1}
	\BP(\beta^+_t = 0) = e^{-\lambda_U t}.
	\end{equation}
	
	\begin{equation}\label{eq:probbetap2}
	\BP(\beta^+_t = 1) =  \frac{\lambda_U }{\lambda - \lambda_U} \big(\exp(-\lambda_U t) - \exp(-\lambda t)\big).
	\end{equation}
	
	For $n\geq 1$ 
	\begin{equation}\label{eq:probbetap3}
	\BP(\beta_t^+ = n+1) =
	\frac{\lambda_U \lambda^n\exp(-\lambda_U t)}{(\lambda - \lambda_U)^{n+1}} 
	-\sum_{k=0}^{n-1} \frac{\lambda_U^2 \lambda^n \exp(-\lambda t)}{(\lambda - \lambda_U)^{n-k+1}} \sum_{j=0}^k \frac{t^j}{j! \lambda^{k-j+1}} - \frac{\lambda_U \exp(-\lambda t)}{\lambda - \lambda_U} \sum_{k=0}^n \frac{t^k \lambda^k }{k!}.
	\end{equation} 
	
	\begin{prop}\label{prop:betapintensity}
		For the stochastic process $\beta^+$, there exist a finite variation stochastic process $\Lambda^+$ such that $\tilde{\beta}^+_t := \beta_t^+ - \Lambda^+_t = \beta_t^+ - \int_0^t \lambda^+_s ds, \tilde{\beta}^+$ is a martingale with respect to the natural filtration $(\calF_t^\beta)_{t \geq 0}$, $\calF_t^\beta = \sigma(\beta_s: s \leq t)$.
	\end{prop}
	
	\proof{Proof}
	By the definition of the process $\beta^+$, it is square integrable. Since $\beta_t^+$ is a nondecreasing process, it is a submartingale. By the Doob--Meyer Decomposition Theorem, there exist a finite variation process $\Lambda^+_t$ such that $\tilde{\beta}_t^+ = \beta_t^+ - \Lambda^+_t, \tilde{\beta}^+$ is a square integrable martingale with respect to the filtration $(\calF_t^\beta)_{ t \geq 0}$. 
	
	Then, by the intensity of Poisson processes $\lambda^+_t = \lim_{h \rightarrow 0} \frac{\BP(\beta_{t+h}^+ - \beta_t^+ = 1)}{h}$ and $\Lambda_t^+ := \int_0^t \lambda^+_s ds$, we have 
	\begin{equation*}
	\begin{split}
	\lim_{h \rightarrow 0} \frac{\BP(\beta_{t+h}^+ - \beta_t^+ = 1)}{h} 
	&= \lim_{h \rightarrow 0} \frac{\BP(\beta_h^+ = 1 | t<T_1)}{h} \one_{\{t<T_1 \}} + \lim_{h \rightarrow 0} \frac{\BP(\beta_{t+h}^+ - \beta_t^+ = 1| T_1 \leq t)}{h} \one_{\{ T_1 \leq t \} }.\\
	\end{split}
	\end{equation*}
	For the second term, when $T_1 < t < t+h$, the process $(\beta_{t+h} - \beta_t)_{t \geq T_1}$ is a Poisson process with the intensity $\lambda$.	By Equation \eqref{eq:probbetap2} and L'Hospital's Rule, the first term is rewrote as
	\begin{equation*}
	\begin{split}
	\lim_{h \rightarrow 0} \frac{\BP(\beta_h^+ = 1 | t<T_1)}{h}  
	& = \frac{\lambda_U }{\lambda - \lambda_U} \lim_{h \rightarrow 0} \big( \lambda \exp(-\lambda h)-\lambda_U  \exp(-\lambda_U h) \big)
	= \lambda_U.
	\end{split}
	\end{equation*}
	Therefore, $\lambda^+_t = \lambda_U \one_{ \{t < T_1 \}} + \lambda \one_{\{T_1 \leq t \}}$, the proposition is proved. 
	
	\endproof

	\begin{thm}\label{thm:betaintensity}
		For an alternating renewal process $\beta$, there exists a finite variation stochastic process $\Lambda^\beta$ such that $ \tilde{\beta}_t := \beta_t - \Lambda^\beta_t = \beta_t - \int_0^t \lambda^\beta_s ds$, $\tilde{\beta}$ is a martingale with respect to the filtration $(\calF_t^\beta)_{t \geq 0}$. 
	\end{thm}
	
	\proof{Proof}
	By the definition of process $\beta$, the integrability is trivial. Since $-\beta^-$ is a Poisson process with parameter $\lambda$, we have that process $\tilde{\beta}^-_t :=\beta_t^- + \lambda t, \tilde{\beta}^-$ is a martingale with respect to the filtration $(\calF_t^\beta)_{t \geq 0}$.  By Proposition \ref{prop:betapintensity}, $\tilde{\beta}^+$ is a martingale with respect to the filtration $(\calF_t^\beta)_{t\geq 0}$. Since $\beta_t = \beta_t^+ + \beta_t^-$, we define $\Lambda_t = \Lambda_t^+ + \lambda t $ and $\lambda^\beta_t = \lambda^+_t + \lambda$, then the expectation is 
	\begin{equation*}
	\begin{split}
	\BE\Big[ \beta_t - \int_0^t \lambda_u^\beta du| \calF_s^\beta\Big] 
	& = \BE\Big[\beta_t^+ - \int_0^t \lambda_u^+ du| \calF_s^\beta\Big] + \BE\Big[ \beta_t^- + \int_0^t  \lambda du | \calF_s^\beta\Big]\\
	& = \beta_s^+ - \int_0^s \lambda_u^+ ds + \beta_s^- + \lambda s\\
	& = \beta_s - \int_0^s \lambda_u^\beta du.
	\end{split}
	\end{equation*}
	So $\tilde{\beta}_t := \beta_t- \Lambda^\beta_t = \beta_t - \int_0^t \lambda_s^\beta ds, \tilde{\beta}$ is a martingale with respect to the filtration $(\calF_t^\beta)_{t \geq 0}$.
	\endproof
	
	\subsection{Jump Counting Processes}\label{sec:poisson_jumpcount}
	In Section \ref{sec:status}, we already define the jump counting processes and discuss several important properties of it. In this section, we will give the proof these propositions.
	
	\begin{prop}\label{Jtilde}
		For the jump counting process $J$, there exists a finite variation stochastic process $\Lambda^J$ such that $\tilde{J}_t := J_t - \Lambda^J_t = J_t - \int_0^t \lambda_s^J ds, \tilde{J}$ is a square integrable martingale with respect to the filtration $(\calF_t^\beta)_{t \geq 0}$.
	\end{prop}
	
	\proof{Proof}
	Since $J = \beta^+ - \beta^-$, we define  $\Lambda_t^J: = \Lambda^+_t + \lambda t, \lambda^J = \lambda^+ + \lambda^-$. Similar to the proof of Theorem \ref{thm:betaintensity}, $\tilde{J}_t := J_t - \Lambda^J_t = J_t - \int_0^t \lambda_s^J ds$ is a martingale with respect to the filtration $(\calF_t^\beta)_{t \geq 0}$ by Proposition \ref{prop:betapintensity}. 
	
	By Proposition \ref{Jsi}, we have $\BE[(\tilde{J}_t)^2] = \BE[(J_t - \int_0^t \lambda_s^J ds)^2] \leq 2\BE[(J_t)^2] + 2\BE[(\int_0^t \lambda^J_s ds)^2] < \infty$. So $\tilde{J}_t$ is a square integrable martingale with respect to the filtration $(\calF^\beta_t)_{t \geq 0}$.
	
	\endproof
	
	We call the stochastic process $\tilde{J}$ as a compensated jump counting process of $J$\index{Compensated jump counting process}.
	
	\subsection{Stochastic Calculus With Respect To Compensated Jump Counting Processes}\label{sec:poisson_stochint}
	
	Since the compensated jump counting process $\tilde{J}$ is a martingale with respect to the filtration $(\calF_t^\beta)_{t \geq 0}$, we can define the stochastic integral with respect to $\tilde{J}$; see \cite[Chapter 4]{metivier1982semimartingales} for details. Let $X$ be a predictable process with respect to the filtration $(\calF_t^\beta)_{t \geq 0}$, and we denote the stochastic integral with respect to the compensated jump counting process $\tilde{J}$ as 
	$
	\int_0^t X_s d \tilde{J}_t.
	$
	
	\begin{prop}[Isometry]
		Let $X$ be a predictable process with respect to the filtration $(\calF_t^\beta)_{t \geq 0}$ and $[\tilde{J}]_t$ be the quadratic variation of the compensated jump counting process $\tilde{J}$. We have 
		$
		\BE\left[\left(\int_0^t X_s d\tilde{J}_s\right)^2\right] = \BE\left[\int_0^t (X_s)^2 d[\tilde{J}]_s\right] = \BE\left[\int_0^t (X_s)^2 \lambda^J_s ds\right].
		$
	\end{prop}
	
	\proof{Proof}
	By Proposition \ref{Jtilde}, we know that $\tilde{J}$ is a square integrable martingale with respect to the filtration $(\calF_t^\beta)_{t \geq 0}$. By Proposition 18.13 in \cite{metivier1982semimartingales}, we have this result.
	\endproof
	
	\subsubsection{The Space of Square Integrable Martingales}
	~\\
	To prove that a square integrable martingale has a representation of stochastic integrals with nonindependent increments processes, we introduce several assumptions and notations at first.
	\begin{ass}\label{orthogonalass}
		Let $W$\index{$W$} be a Brownian motion, $\beta$ be an alternating renewal process, $\varpi^I, \varpi^C$\index{$\varpi^i$} be two compensated processes with a single exponential distributed jump, which are independent and strongly orthogonal.
	\end{ass}
	
	\begin{notation}
		\begin{itemize}
			\item 	$\calF_t = \sigma(W_s, \beta_s, \varpi_s^I, \varpi_s^C : s\leq t).$\index{$\calF_t$}
			
			\item $\mathscr{H}^{\beta, 2} = \{ X | X$ is $\mathscr B([0, t]) \otimes \calF_t^\beta ~\text{predictable process with }  \| X\|_{\mathscr{H}_t^2} <\infty, \text{ for } \forall t \leq T \},$ where $\| \cdot\|_{\mathscr{H}_T^2} = \BE[\int_0^t X_s^2 ds]$.\index{$\mathscr{H}^{\beta, 2}$}
			
			\item $\mathscr{H}^2 = \{ (X, X^I, X^C, X^\beta) | X, X^I, X^C, X^\beta$ are $\mathscr B([0,t]) \otimes \calF_t ~\text{predictable process with }  \|X\|_{\mathscr{H}_T^2} < \infty, \|X^I\|_{\mathscr{H}_T^2}< \infty, \|X^C\|_{\mathscr{H}_T^2} < \infty \text{ and } \|X^\beta\|_{\mathscr{H}_T^2} < \infty, \text{ for } \forall t \leq T\}.$\index{$\mathscr{H}^2$}

			\item $\mathscr M^\beta = \{ M| M \text{ is a } (\calF_t^\beta)_{t \geq 0} \text{ martingale with } \sup_{t\leq T} \BE[M_t^2] < \infty,\text{ for } \forall t \leq T \}$.\index{$\mathscr M^\beta$} 
			
			\item $\mathscr M = \{ M| M \text{ is a } (\calF_t)_{t \geq 0} \text{ martingale with } \sup_{t\leq T} \BE[M_t^2] < \infty, \text{ for } \forall t \leq T\}$.\index{$\mathscr M$}
			
			\item $\mathscr{M}^{\beta, *}_T = \{M_T| M_T \text{ is a } \calF_T^\beta \text{ measurable random variable with }M_T: = I^\beta_T(X) = \int_0^T X_s d\tilde{J}_s,\newline \text{ for } \sup_{t\leq T} \BE[M_t^2] < \infty, X \in \mathscr H^{\beta,2} \}$.\index{$\mathscr{M}^{\beta, *}_T$}
			
			\item $\mathscr{M}^{\beta, *} = \{M| M  \in \mathscr{M}^\beta \text{ and } M_t: = I^\beta_t(X) = \int_0^t X_s d\tilde{J}_s \text{ with } X \in \mathscr H^{\beta,2},\text{ for } \forall t \leq T\}$.\index{$\mathscr{M}^{\beta, *}$}
			
			\item $\mathscr{M}^*_T = \{M_T | M_T \text{ is a } \calF_T \text{ measurable random variable with } M_T =: I_T(X) = \int_0^T X_s dW_s + \int_0^T X^I_s d \varpi_t^I + \int_0^T X^C_s d\varpi_t^C + \int_0^T X^\beta_s d\tilde{J}_s, \text{ for } \sup_{t\leq T} \BE[M_t^2] < \infty, (X, X^I, X^C, X^\beta) \in \mathscr H_T^2 \}$.\index{$\mathscr{M}^*_T$}
			
			\item $\mathscr{M}^* = \{M |M \in \mathscr{M} \text{ and } M_t =: I_t(X) = \int_0^t X_s dW_s + \int_0^t X^I_s d \varpi_s^I + \int_0^t X^C_s d\varpi_s^C + \int_0^t X^\beta_s d\tilde{J}_s \text{ with }\newline (X, X^I, X^C, X^\beta) \in \mathscr H^2,\text{ for } \forall t \leq T \}$.\index{$\mathscr{M}^*$}
			
		\end{itemize}
	\end{notation}
	
	\begin{prop}\label{betam2com}
		Given $(\Omega, \calF,  (\calF_t^\beta)_{t \geq 0}, \BP)$, $(\mathscr{M}^{\beta, *}, \|~\|)$ is a Banach space with the norm $\|\cdot\|^2 = \BE[M_t^2]$ for $M\in \mathscr{M}^{\beta, *}$.
	\end{prop}
	
	\proof{Proof}

	I. We prove first that $(\mathscr{M}^{\beta, *}, \|~\|)$ is a vector space. Let $M_t^{(1)}, M_t^{(2)} \in \mathscr{M}^{\beta, *}$, then we have two predictable square integrable processes $X^{(1)}$ and $X^{(2)}$ such that $M_t^{(1)} = \int_0^t X_s^{(1)} d\tilde{J}_s, M_t^{(2)} = \int_0^t X_s^{(2)} d\tilde{J}_s$. Let $\mu_1, \mu_2 \in \RR$ and $X = \mu_1 X^{(1)} + \mu_2 X^{(2)}$, then $X$ is still predictable and square integrable. So $\mu_1 M_t^{(1)} + \mu_2 M_t^{(2)} = \int_0^t X_s d\tilde{J}_s \in \mathscr{M}^{\beta,*}$.
	
	II.
	Let $M_t^{(n)} \in \mathscr{M}^{\beta, *}$ be a Cauchy sequence. Since this vector space is a topological normed space, the limit of $M_t^{(n)}$ exists, denoted by $M_t = \lim_{n\rightarrow \infty} M_t^{(n)}$. Then we need to prove that $M_t \in \mathscr{M}^{\beta,*}$. Since $M_t^{(n)} \in \mathscr{M}^{\beta,*}$, there exists a sequence of adapted square integrable processes $X^{(n)}$ such that $M_t^{(n)} = I_t^\beta(X^{(n)}) = \int_0^t X^{(n)}_s d\tilde{J}_s$.
	
	i) Let $X = \lim_{ n \rightarrow \infty} X^{(n)}$, we need to prove it exist and $X$ is predictable and square integrable. Since $M_t^{(n)}$ is a Cauchy sequence, for any given $\epsilon > 0$, there exists $N \in \BN$ such that $\| M_t^{(n)} - M_t^{(m)} \| < \epsilon$ for any $n,m > N$. Since $M_t^{(n)}, M_t^{(m)} \in \mathscr{M}^{\beta, *}$, there exist $X^{(n)}, X^{(m)}$ such that $M_t^{(n)} = \int_0^t X_s^{(n)} d\tilde{J}_s, M_t^{(m)} = \int_0^t X_s^{(m)} d\tilde{J}_s $. Based on the definition of $\mathscr{H}^{\beta, 2}$ and the isometry property, we have 
	\begin{equation*}
	\begin{split}
	\|X^{(n)} - X^{(m)}\|_{\mathscr{H}_t^2} 
	\leq & \frac{1}{\lambda} \BE\left[ \int_0^t |X_s^{(n)} - X_s^{(m)}|^2 \lambda_s^J ds \right]\\
	= & \frac{1}{\lambda} \BE\left[ \int_0^t |X_s^{(n)} - X_s^{(m)}|^2 d[\tilde{J}]_s\right]\\
	= & \frac{1}{\lambda} \BE\Big[ \Big(\int_0^t |X_s^{(n)} - X_s^{(m)}| d\tilde{J}_s\Big)^2 \Big] \\
	= & \frac{1}{\lambda} \|M^{(n)} - M^{(m)}\|^2_t\\
	< &\epsilon.
	\end{split}
	\end{equation*}
	
	So the sequence $X^{(n)}$ is a Cauchy sequence. Since $\mathscr{H}^{\beta,2}$ is complete, the limit $X = \lim_{ n \rightarrow \infty} X^{(n)}$ exists. By the definition of $X$, the predictability and square integrability properties are trivial. 
	
	ii)  For any $\epsilon > 0$, there exists $N \in \BN$ such that $\|M_t^{(n)} - M_t\|^2  < \epsilon$ for any $n > N$. We need to prove the square integrability. For any $n > N$ be given, we have $\|M\|^2 \leq (\|M_t - M_t^{(n)}\| + \|M_t^{(n)}\|)^2 \leq (\epsilon + \|M_t^{(n)}\|)^2 < \infty$, which proved the square integrability. By the square integrability, $M_t = \lim_{n\rightarrow \infty} M_t^{(n)} = \lim_{n\rightarrow \infty} \int_0^t X_s^{(n)} d\tilde{J}_s = \int_0^t \lim_{ n \rightarrow \infty} X_s^{(n)} d\tilde{J}_s = \int_0^t X_s d\tilde{J}_s$, which proved $M_t \in \mathscr{M}^{\beta, *}.$
	
	So any Cauchy sequence $M_t^{(n)} \in \mathscr M^{\beta,*}$ is convergent in the space $(\mathscr{M}^{\beta, *}, \|~\|)$, which means $(\mathscr{M}^{\beta,*}, \|~\|)$ is close.
	
	Overall, $(\mathscr{M}^{\beta,*}, \|~\|)$ is a Banach space.
	\endproof
	
	\begin{rem}
		By the square integrability in this Banach space, it is also a Hilbert space for a inner produce $<M_t^{1},M_t^{2}> =\BE[M_t^{1}M_t^{2}]$.
	\end{rem}
	
	\begin{prop}\label{betaITXmart}
		$\mathscr{M}^{\beta,*}_t \in \mathscr{M}^{\beta}$, which means for any $M_t:=I_t^\beta(X) \in \mathscr{M}^{\beta,*}$, $I^\beta(X)$ is a square integrable martingale with respect to the filtration $(\calF_t^\beta)_{t\geq 0}$.
	\end{prop}
	
	\proof{Proof}
	I. By Proposition \ref{betam2com}, the square integrability of $I_t^\beta(X)$ follows the square integrability of $X$.
	
	II. For a given $t \geq 0$, we need to prove the martingale property of $I_t^\beta(X)$. By the definition of process $\tilde{J}$, We have
	
	\begin{align}\label{eq:Icondprob}
	\BE[I^\beta_T(X) | \calF_t^\beta] 
	& = \BE\Big[\int_0^t X_s d\tilde{J}_s|\calF_t^\beta\Big] + \BE\Big[\int_t^T X_s d\tilde{J}_s | \calF_t^\beta\Big]
	= I^\beta_t(X) + \BE\Big[\int_t^T X_s d\tilde{J}_s | \calF_t^\beta\Big]
	\end{align}
	
	Next, We show that the second term in the above equation is equal 0 in two steps.
	
	i) Assume that $X$ is an elementary process, i.e. $X = \sum_{i=1}^{N} a_{t_i}\one_{\{t_i \leq s < t_{i+1}\}} $ with $a_{t_i}$ are $\calF_{t_i}^\beta$ measurable square integrable random variables and $0 = t_1 < \cdots < t_{k-1} < t < t_k < \cdots < t_N = T$. Then the second term in Equation \eqref{eq:Icondprob} becomes
	\begin{align}\label{eq:Jtilde}
	\BE\Big[\int_t^T X_s d\tilde{J}_s|\calF_t^\beta\Big] 
	&= \sum_{i=1}^{N-1} \BE\Big[  a_{t_i}\int_t^T \one_{\{t_i \leq s < t_{i+1}\}} d\tilde{J}_s|\calF_t^\beta\Big]\nonumber\\
	& =  \BE\Big[\BE[  a_{t_{k-1}}(\tilde{J}_{t_{k}} - \tilde{J}_{t})|\calF_{t_i}^\beta]\Big|\calF_t^\beta\Big] + \sum_{i=k}^{N-1}\BE\Big[\BE[  a_{t_i}(\tilde{J}_{t_{i+1}} - \tilde{J}_{t_i})|\calF_{t_i}^\beta]\Big|\calF_t^\beta\Big]\nonumber\\
	& = \BE\Big[a_{t_{k-1}} \BE[ \tilde{J}_{t_{k}} - \tilde{J}_{t}|\calF_{t_i}^\beta]\Big|\calF_t^\beta\Big] + \sum_{i=k}^{N-1}\BE\Big[ a_{t_i} \BE[  \tilde{J}_{t_{i+1}} - \tilde{J}_{t_i}|\calF_{t_i}^\beta]\Big|\calF_t^\beta\Big].
	\end{align}
	By the definition of $\tilde{J}_t = \sum_{s \leq t}\one_{\{\beta_s - \beta_{s-} \neq 0\}}(s) - \int_0^t \lambda_s^J ds$, we have 
	\begin{equation*}
	\begin{split}
	\tilde{J}_{t_{i+1}} -\tilde{J}_{t_i} & = \sum_{s \leq t_{i+1}}\one_{\{\beta_s - \beta_{s-} \neq 0\}}(s) - \int_0^{t_{i+1}} \lambda_s^J ds - \Big( \sum_{s \leq t_{i}}\one_{\{\beta_s - \beta_{s-} \neq 0\}}(s) - \int_0^{t_{i}} \lambda_s^J ds \Big)\\
	& = \sum_{t_i < s \leq t_{i+1}}\one_{\{\beta_s - \beta_{s-} \neq 0\}}(s) - \int_{t_i}^{t_{i+1}} \lambda_s^J ds.
	\end{split}
	\end{equation*}
	Let $s' = s-t_i$ for $s \geq t_i$, we have $s = s' + t_i$, then
	\begin{equation*}
	\begin{split}
	\tilde{J}_{t_{i+1}} -\tilde{J}_{t_i} & = \sum_{t_i \leq s' + t_i \leq t_{i+1}} \one_{\{\beta_{s'+t_i} - \beta_{(s'+t_i)-}\}}(s') - \int_{t_i}^{t_{i+1}} \lambda_{s'+t_i}^J d(s'+t_i)\\
	& = \sum_{0 < s' \leq t_{i+1} - t_i} \one_{\{\beta_{s'+t_i} - \beta_{(s'+t_i)-} \neq 0\}}(s') - \int_{0}^{t_{i+1}-t_i} \lambda_{s'+t_i}^J d(s').\\
	\end{split}
	\end{equation*}
	Let $\beta'_t = \beta_{t'+t_i}$ and $\lambda^{J'}_t = \lambda^J_{t'+t_i}$, we have
	$$
	\tilde{J}_{t_{i+1}} -\tilde{J}_{t_i} = \sum_{0\leq s \leq t} \one_{\{ \beta_s' - \beta_{s-}' \neq 0\} }(s) - \int_0^t \lambda_s^{J'} ds,
	$$
	denoted by $\tilde{J}'_t$.  By the definition of compensated jump counting processes, $\tilde{J}'$ is a compensated jump counting process by the memoryless property of exponential processes. When $\beta_{t_i} = 0$, $\tilde{J}'$ has same distribution as $\tilde{J}$. When $\beta_{t_i} = 1$, $\tilde{J}'$ is a compensated jump counting process starting with the initial value 1. In summary, we have $\BE[\tilde{J}'_t \one_{\{\beta_{t_i} = 0\}}] = 0$ and $\BE[\tilde{J}'_t \one_{\{ \beta_{t_i} = 1 \}}] = 0$ for $t \geq t_i$. By the Markov property of a Poisson process, we have
	\begin{equation*}
	\begin{split}
	\BE[\tilde{J}_{t_{i+1}} - \tilde{J}_{t_i}| \calF_{t_i}^\beta] & = \BE[\tilde{J}'_{t_{i+1}}| \sigma(\beta_{t_i})]\\
	& = \frac{\BE[\tilde{J}'_{t_{i+1}}\one_{\{ \beta_{t_i}=1 \}}]}{\BP(\beta_{t_i}=1)}\one_{\{ \beta_{t_i}=1 \}} +  \frac{\BE[\tilde{J}'_{t_{i+1}}\one_{\{ \beta_{t_i}=0 \}}]}{\BP(\beta_{t_i} = 0 )}\one_{\{ \beta_{t_i}=0 \}}\\
	& = 0.\\
	\end{split}
	\end{equation*}
	Similarly, the first term in Equation \eqref{eq:Jtilde} becomes $\BE\Big[a_{t_{k-1}} \BE[ \tilde{J}_{t_{k}} - \tilde{J}_{t}|\calF_{t_i}^\beta]\Big|\calF_t^\beta\Big]  = 0$. Thus, the second term in Equation \eqref{eq:Icondprob} becomes
	$$
	\BE\Big[a_{t_{k-1}} \cdot 0\Big|\calF_t^\beta\Big] + \sum_{i=k}^N\BE\Big[ a_{t_i} \cdot 0\Big|\calF_t^\beta\Big] =  0.
	$$
	Therefore, $\BE[I_T^\beta(X)| \calF_t^\beta] = I_t^\beta(X)$. Since $t$ is any arbitrage time, we proved the martingale property of $I^\beta(X)$ when $X$ is an elementary process.
	
	ii) For any predictable square integrable stochastic process $X$, there exists a sequence of square integrable elementary processes $X^{(n)} \rightarrow X$, as $n \rightarrow \infty$. By the square integrability, the second term in Equation \eqref{eq:Icondprob} becomes 
	\begin{equation*}
	\begin{split}
	\BE\Big[\int_t^T X_s d\tilde{J}_s|\calF_t^\beta\Big] &= \BE\Big[ \int_t^T \lim_{n\rightarrow \infty} X_s^{(n)}d\tilde{J}_s|\calF_t^\beta \Big] \\
	& =  \lim_{n\rightarrow \infty}\BE\Big[ \int_t^T  X_s^{(n)}d\tilde{J}_s|\calF_t^\beta \Big] \\
	& =  \lim_{n\rightarrow \infty} 0\\
	& = 0.
	\end{split}
	\end{equation*}
	So we proved the martingale property of $I^\beta(X)$.
	\endproof
	
	Assume that the stochastic processes $W, \beta, \varpi^I, \varpi^C$ are independent and strongly orthogonal, Proposition \ref{betam2com} and Proposition \ref{betaITXmart} can be extended to the martingale $M$ with respect to the natural filtration $(\calF_t)_{t \geq 0}$. 
	
	\begin{prop}\label{m2com}
		Given $(\Omega, \calF, (\calF_t)_{t \geq 0}, \BP)$, $(\mathscr{M}^*, \|~\|)$ is a Banach space with the norm $\|\cdot\|^2 = \BE[M_t^2]$ for $M_t \in M^*$.
	\end{prop}
	
	\proof{Proof}
	I. First we prove that $(\mathscr{M}^*, \|~\|)$ is a vector space.
	Let $M_t^1, M_t^2 \in \mathscr M^*$ and $\mu_1, \mu_2 \in \RR$. By the linearity of stochastic integrals, we have 
	\begin{equation*}
	\begin{split}
	\mu_1M^1_t +\mu_2 M^2_t = &\int_0^t (\mu_1 X^1_s + \mu X^2_s ) dW_s + \int_0^t (\mu_1 X^{1,I}_s + \mu_2 X^{2,I}_s) d\varpi_s^I\\
	& \quad + \int_0^t (\mu_1 X^{1,C}_s + \mu_2 X^{2,C}_s) d\varpi_s^C + \int_0^t (\mu_1 X^{1,\beta}_s + \mu_2 X^{2,\beta}_s) d\tilde{J}_s \\
	& \in \mathscr M_t^*.
	\end{split}
	\end{equation*}
	
	II. Since $(\calF_t)_{t \geq 0}, \BP)$ is a normed space, we want to prove the completeness of this space.
	By the definition of $\|\cdot\|^2$, Proposition \ref{betam2com} and the strongly orthogonality of the processes $W, \beta, \varpi^I, \varpi^C$, we have the completeness of $(\mathscr M^*, \| ~\|)$.
	
	Overall, we proved that  $(\mathscr{M}^*, \|~\|)$ is a Banach space.
	\endproof

	\begin{prop}\label{ITXmart}
		$\mathscr{M}^* \in \mathscr{M}$, which means for any $M_t = I_t(X) \in \mathscr{M}^*$, $I(X)$ is a square integrable martingale with respect to the filtration $(\calF_t)_{t \geq 0}$.
	\end{prop}
	
	\proof{Proof}
	I. For the square integrability, since $(X, X^I,X^C, X^\beta) \in \mathscr H^2$, we have this property.
	
	II. Then we want to prove the martingale property of $I(X)$.
	By the linearity of the expectation, independence and Proposition \ref{betaITXmart}, we have 
	\begin{equation*}
	\begin{split}
	\BE\Big[I_T(X) |\calF_t\Big] 
	& = \BE\Big[\int_0^T X_s dW_s | \calF^W_t\Big] + \BE\Big[ \int_0^T X^I_s d\varpi_s^I | \calF^I_t\Big] + \BE\Big[ \int_0^T X^C_s d\varpi_s^C | \calF^C_t\Big] + \BE\Big[ \int_0^T X^\beta_s d\tilde{J}_s | \calF^\beta_t\Big]\\
	& = \int_0^t X_s dW_s + \int_0^t X^I_s d\varpi_s^I + \int_0^t X^C_s d\varpi_s^C + \int_0^t X^\beta_s d\tilde{J}_s \\
	& = I_t(X).
	\end{split}
	\end{equation*}
	Thus, $\mathscr{M}^*$ is a subspace of $\mathscr{M}$.
	\endproof
	
	\subsubsection{Martingale Decomposition Theorem}
	~\\
	In this section, we will prove the existence and uniqueness of a decomposition of a square integrable martingale. Any square integrable martingale with respect to the filtration $(\calF_t)_{t\geq 0}$ can be write in a form as the sum of stochastic integrals with respect to $W, \tilde{J}, \varpi^I, \varpi^C$ and a orthogonal term. Similar to Proposition 4.1 in \cite{kunita1967square}, we will prove a decomposition for a random variable $M_T$ and then extend the decomposition ot any martingale $M \in M_t$.
	
	\begin{prop}\label{kunita}
		Let $M_T$ be an $\calF_T^\beta$ measurable random variable with $\sup_{t\leq T} \BE[M_t^2] < \infty$, $(\mathscr M_T^*)^\perp$ be the orthogonal space with respect to $\mathscr M_T^*$. Then, there exists a unique pair $I_T(X) \in \mathscr M_T^*$ and $Y \in (\mathscr M_T^*)^\perp$ such that $M_T = I_T(X) + Y_T$.
	\end{prop}
	
	\proof{Proof}
	I. First we want to prove the existence of the decomposition $M_T = I_T(X) + Y_T$.
	Because $\mathscr M_T^*$ is complete subspace of $\mathscr M_T$, there exists an orthogonal space of $\mathscr M_T^*$ in $\mathscr M_T$, denoted as $(\mathscr M_T^*)^\perp$.
	
	Let $I_T(X) = \textit{proj}_{\mathscr M^*_T} M_T, Y = M_T - I_T(X)$. We have $M_T = I_T(X) + Y$ with
	$$
	\|I_T(X)Y\|^2 = \|I_T(X)(M_T-I_T(X))\|^2 = \|(\textit{proj}_{\mathscr M^*_T} M_T)^2 - (\textit{proj}_{\mathscr M^*_T} M_T)^2\|^2 = 0.
	$$
	So we have $Y \in (\mathscr M^*_T)^\perp$. We proved the existence of the decomposition.
	
	II. Then, we want to prove the uniqueness of the decomposition.
	Assume that there exists two decompositions, $ I_T^1(X), I_T^2(X)\in \mathscr M_T^*$ and $Y_1, Y_2 \in (\mathscr M_T^*)^\perp$ such that $M_T = I_T^1(X) + Y_1  = I_T^2(X) + Y_2 $. Thus, $I_T^1(X) - I_T^2(X) = Y_2 - Y_1 $. Since $\mathscr M^*_T$ and $(\mathscr M^*_T)^\perp$ are linear spaces and $I_T^1(X) - I_T^2(X)\in \mathscr M^*_T$ and $Y_2 - Y_1 \in (\mathscr M^*_T)^\perp$, we have $I_T^1(X) - I_T^2(X) = Y_2 - Y_1 \in \mathscr M_T^* \cap (\mathscr M^*_T)^\perp$. By the property of an orthogonal space, $I_T^1(X) - I_T^2(X) = Y_2 - Y_1 = 0$. Therefore, we get $I_T^1(X) = I_T^2(X) \in \mathscr M^*_T$ and $Y_1 = Y_2 \in (\mathscr M^*_T)^\perp$. So, we proved the uniqueness.
	\endproof
	
	By the unique decomposition of $M_T \in \mathscr M_T$, we have the following decomposition of a martingale $M \in \mathscr M$.
	
	\begin{thm}\label{martdecom}\index{Martingale decomposition theorem}
		Let $M \in \mathscr M_t$. Then, there exists a unique decomposition $M = I(X)+Y$ within the space $(\mathscr M, \|~\|)$, where $I(X) \in \mathscr{M}^*$ and $Y \in (\mathscr{M}^*)^\perp$.
	\end{thm}
	
	\proof{Proof}
	
	I. We want to prove the existence of a decomposition of $M \in \mathscr M$.
	By Proposition \ref{kunita}, for any $M_T \in \mathscr M_T$, we have $M_T = Y_T + I_T(X)$, where 
	$
	I_T(X) =\int_0^T X_s dW_s + \int_0^T X^I_s d \varpi_t^I + \int_0^T X^C_s d\varpi_t^C + \int_0^T X^\beta_s d\tilde{J}_s \in M^*_T,
	$
	and $Y_T \in (M^*_T)^{\perp}$. 
	
	Since $M$ is a martingale, by Proposition \ref{ITXmart}, we have
	$
	M_t = \BE[M_T | \calF_t] = \BE[Y_T + I_T(X) |\calF_t] = \BE[Y_T | \calF_t] + I_t(X).
	$
	Define a stochastic process $Y$ by $Y_t: = \BE[Y_T |\calF_t]$. So we have a decomposition of $M_t = Y_t + I_t(X)$. Then, we want to prove that $Y \in (\mathscr M^*)^\perp$. Since $I(X) \in \mathscr M^*$, we denote $I_t(X) = \int_0^t X_s dW_s + \int_0^t X^I_s d \varpi_t^I + \int_0^t X^C_s d\varpi_t^C + \int_0^t X^\beta_s d\tilde{J}_s \in \mathscr M^*$. Then
	\begin{equation*}
	\begin{split}
	\BE[Y_t I_t(X)] 
	&= \BE\Big[\BE[Y_T | \calF_t] \Big( \int_0^t X_s dW_s + \int_0^t X^I_s d \varpi_s^I + \int_0^t X^C_s d\varpi_s^C + \int_0^t X^\beta_s d\tilde{J}_s \Big) \Big] 
	\\
	& = \BE\Big[\BE\Big[Y_T \Big( \int_0^t X_s dW_s + \int_0^t X^I_s d \varpi_s^I + \int_0^t X^C_s d\varpi_s^C + \int_0^t X^\beta_s d\tilde{J}_s \Big) |\calF_t\Big]\Big]
	\\
	&= \BE\Big[Y_T \Big( \int_0^t X_s dW_s + \int_0^t X^I_s d \varpi_s^I + \int_0^t X^C_s d\varpi_s^C + \int_0^t X^\beta_s d\tilde{J}_s \Big) \Big].
	\end{split}
	\end{equation*}
	
	Define $X'_s = \one_{ \{  s \leq t \}} X_s, X^{'I}_s = \one_{ \{  s \leq t \}} X^I_s, X^{'C}_s = \one_{ \{  s \leq t \}} X^C_s, X^{'\beta}_s = \one_{ \{  s \leq t \}} X^\beta_s,$ $\forall s \in [0,T],$ $(X_s', X^{'I}_s, X^{'C}_s, X^{'\beta}_s) \in \mathscr H^2$, we have $I_t(X) = I_T(X') \in \mathscr M_T^*$. Hence
	$$
	\BE\Big[Y_t I_t(X)\Big] 
	= \BE\Big[Y_T \Big( \int_0^T X'_s dW_s + \int_0^T X^{'I}_s d \varpi_s^I + \int_0^T X^{'C}_s d\varpi_s^C + \int_0^T X^{'\beta}_s d\tilde{J}_s \Big) \Big] 
	= 0.
	$$
	Since $Y \perp I(X)$ for any $I(X) \in \mathscr M^*$, we proved $Y \in (\mathscr M^*)^\perp$.
	
	II. Next, we want to prove the uniqueness of this decomposition.
	For any given $M \in \mathscr M$, we assume that $I^1(X), I^2(X) \in \mathscr M^*$ and $Y^1, Y^2 \in (\mathscr M^*)^\perp$ such that $M_t = Y^1_t + I^1_t(X) = Y^2_t + I^2_t(X)$, for $0\leq t \leq T < \infty$. Then $0 = M_t - M_t$ becomes
	{\small
		$$
		0 =Y^1_t - Y^2_t  +\int_0^t (X^1_s - X^2_s) dW_s + \int_0^t (X^{I,1}_s - X^{I,2}_s) d \varpi_s^I + \int_0^t (X^{C,1}_s - X^{C,2}_s) d\varpi_s^C + \int_0^t (X^{\beta,1}_s - X^{\beta,2}_s) d\tilde{J}_s ,
		$$
	}
	which means that
	$$
	Y^2_t - Y^1_t = \int_0^t (X^1_s - X^2_s) dW_s + \int_0^t (X^{I,1}_s - X^{I,2}_s) d \varpi_s^I + \int_0^t (X^{C,1}_s - X^{C,2}_s) d\varpi_s^C + \int_0^t (X^{\beta,1}_s - X^{\beta,2}_s) d\tilde{J}_s \in \mathscr M^*.
	$$
	Since $Y^2 - Y^1 \in (\mathscr M^*)^\perp \cap \mathscr M^*$, we have $Y^2 - Y^1 = 0 $ within the space $(\mathscr M, \| ~\|)$. And $I^1(X) - I^2(X) = 0$. So the decomposition is unique in the space $(\mathscr M, \|~\|)$. 
	\endproof 
	
	\section{Backward Stochastic Differential Equations}\label{chap:bsde}
	In this appendix, we will define a general backward stochastic differential equation (BSDE), including a stochastic integral with respect to the process $\tilde{J}$. Because $\tilde{J}$ does not have the independent increments property, we need to apply Theorem \ref{martdecom} and some results of a fixed point problem to prove the existence and uniqueness of the solutions. Then we will rewrite the original BSDEs with the filtration $(\calF_t)_{t \geq 0}$ to a smaller filtration and show their equivalence.
	
	For convenience, we define several notations here.
	\begin{notation}
		\begin{itemize}
			\item $\HH^2 = \{X|X: \Omega \times [0,T] \rightarrow \RR \text{ is a predictable process with } \BE\big[\int_0^T |X_s|^2 ds\big] < \infty\}$.\index{$\HH^2$}
			
			\item $\BS^2 = \{ X| X: \Omega \times [0,T] \rightarrow \RR \text{ is a $c\grave{a}dl\grave{a}g$ adapted processes with } \BE\big[\sup_{s \in [0,T]} |X_s|^2\big] < \infty\}$.\index{$\BS^2$}
			
			\item $\BM^2 = \{ M| M  \text{is a martingale with respect to }(\calF_t)_{t \geq 0} \text{ in } \BS^2 \}$.\index{$\BM^2$}
		\end{itemize}
	\end{notation}
	
	\subsection{Construction of the General BSDEs}\label{sec:bsde_construct}
	
	Let $N^I, N^C$\index{$N^i$} be two Poisson processes  with parameters $\lambda^I, \lambda^C$\index{$\lambda^i$}. Define $\tau^I = \inf\{t : N^I_t = 1\}, \tau^C = \inf\{t: N^C_t = 1\}$\index{$\tau^i$} and $\tau = \tau^I \wedge \tau^C$\index{$\tau$}. Let $\tilde{J}$ a corresponding compensated jump counting process with parameter $\lambda^J$.  Assume that $W, \tilde{J}, \varpi^I$ and $\varpi^C$\index{$\varpi^i$} are independent and strongly orthogonal. In this section, we study a general BSDE on the filtered probability space $(\Omega, \calF, (\calF_t)_{t\geq 0}, \BP)$, where $\calF_t = \sigma (W_s, \beta_s, N^I_s, N^C_s: s \leq t)$\index{$\calF_t$} as following form.
	
	\begin{equation}\label{generalbsde}\index{General BSDE}
	\left\{\begin{aligned}
	dV_t &=  f(\omega, t, V_{t-}, Z_t, Z^I_t, Z^C_t, Z^\beta_t)dt -  Z_t dW_t
	-  Z^I_t d\varpi^I_t -  Z^C_t d\varpi^C_t -  Z^\beta_t d \tilde{J}_t,
	\\
	V_T &= \theta,
	\end{aligned}
	\right.
	\end{equation}
	where $ 0 \leq t \leq T < \infty, (Z, Z^I, Z^C, Z^\beta) \in \mathscr H^2$.
	
	To simplify notations, we define a martingale $M$ of a specific form and a generator $F_t(V,M)$ \index{$F_t(V,M)$}as follows.
	
	\begin{dfn}
		Define $ M_t := \int_0^t Z_s dW_s + \int_0^t Z^I_s d \varpi_t^I + \int_0^t Z^C_s d\varpi_t^C + \int_0^t Z^\beta_s d\tilde{J}_s $, for $(Z, Z^I, Z^C, Z^\beta) \in \mathscr{H}^2 $ and  $0\leq t \leq T < \infty$.
	\end{dfn}	
	
	\begin{prop}
		$ M$ is a martingale with respect to $(\calF_t)_{t \geq 0}$.
	\end{prop}
	
	\proof{Proof}
	Since $(Z, Z^I, Z^C, Z^\beta) \in \mathscr H^2$, we have that $Z_t, Z^I_t, Z^C_t, Z^\beta_t$ are $\mathscr B([0,t]) \otimes \calF_t$ predictable and square integrable. By Proposition \ref{betaITXmart}, we have that $M$ is a martingale.
	\endproof
	
	\begin{dfn}
		Define a generator as a function $F_t(V, M) : \mathscr{H}^2 \times \mathscr{M}^* \rightarrow \BS^2$ with $F_t(V, M) =  \int_0^t f(\omega, s, V, Z, Z^I, Z^C, Z^\beta) ds$.
	\end{dfn}

	For general BSDE \eqref{generalbsde}, we have the following assumptions:
	\begin{ass}\label{existuniqueass}
		~\\\vspace{-0.5cm}
		\begin{enumerate}[(i)]
			\item (Lipschitz Condition)\index{Lipschitz Condition} The generator $f: \Omega \times [0,T] \times \RR^5 \rightarrow \RR, (\omega, t, v, z, z^I, z^C, z^\beta)\mapsto (\omega, t)$ is predictable and Lipschitz continuous in $v, z, z^I, z^C, z^\beta$, i.e. for  $(v_1, z_1, z^I_1, z^C_1, z^\beta_1), (v_2, z_2, z^I_2, z^C_2, z^\beta_2) \in \RR^5$, we have
			$$
			\begin{aligned}
			&|f(\omega, t, v_1, z_1, z^I_1, z^C_1, z^\beta_1) - f(\omega, t, v_2, z_2, z^I_2, z^C_2, z^\beta_2)|\\
			< &\frac{1}{5T}\left(\frac{1}{\sqrt{T}}|v_1 - v_2| + |z_1 - z_2| + \sqrt{\lambda^I}|z^I_1 - z^I_2| + \sqrt{\lambda^C}|z^C_1 - z^C_2| + \sqrt{\lambda^J}|z^\beta_1 - z^\beta_2|\right).
			\end{aligned}
			$$
			\item (Terminal Condition)\index{Terminal Condition} The terminal value satisfies $\theta \in L^2 (\Omega, \calF_T, \BP).$
			
			\item (Integrability Condition)\index{Integrability Condition} $f(\omega, t, 0,0,0,0,0) \in \HH^2.$
		\end{enumerate}
	\end{ass}
	
	\begin{prop}\label{lipF}	
		The generator $F_t(V,M)$ satisfies the following inequality
		$$
		\|F_t(V_1, M_1) - F_t(V_2, M_2)\|_{\BS^2} < \frac{1}{5}\Big(\|V_1 - V_2\|_{\BS^2} + \|Z_1 - Z_2\|_{\BS^2} + \|Z^I_1 - Z^I_2\|_{\BS^2} + \|Z^C_1 - Z^C_2\|_{\BS^2} + \|Z^\beta_1 - Z^\beta_2\|_{\BS^2}\Big).
		$$
	\end{prop}
	
	\proof{Proof}
	Since the function $f$ satisfies the Lipschitz condition in Assumptions \ref{existuniqueass} and $W, \tilde{J}, \varpi^I, \varpi^C$ are orthogonal, we have 
	\begin{equation*}
	\begin{split}
	&\|F_t(V_1, M_1) - F_t(V_2, M_2)\|_{\BS^2} \\
	< &\BE \Big[ \sup_{0 \leq t \leq T}\int_0^t \frac{1}{5} \Big(\frac{1}{T}|V_1 -V_2|^2 + |Z_1 - Z_2|^2 + \lambda^I | Z^I_1 - Z^I_2|^2 + \lambda^C | Z^C_1 - Z^C_2|^2 + \lambda^J| Z^\beta_1 - Z^\beta_2|^2\Big)  ds \Big]
	\\
	\leq & \frac{1}{5T} \BE \Big[ \sup_{0 \leq t \leq T} \int_0^t |V_1 -V_2|^2 ds \Big] 
	\\
	&\quad + \frac{1}{5}\BE \Big[ \sup_{0 \leq t \leq T} \Big(\int_0^t |Z_1 - Z_2|^2 ds  +\int_0^t \lambda^I | Z^I_1 - Z^I_2|^2 ds  +\int_0^t \lambda^C | Z^C_1 - Z^C_2|^2 ds  +\int_0^t \lambda^J| Z^\beta_1 - Z^\beta_2|^2ds \Big) \Big].
	\\
	\end{split}
	\end{equation*}
	By the properties of isometry and orthogonality, the second term is
	\small{
		$$
		\frac{1}{5}\BE \Big[ \sup_{0 \leq t \leq T} \Big(\int_0^t |Z_1 - Z_2|^2 ds  +\int_0^t \lambda^I | Z^I_1 - Z^I_2|^2 ds  +\int_0^t \lambda^C | Z^C_1 - Z^C_2|^2 ds  +\int_0^t \lambda^J| Z^\beta_1 - Z^\beta_2|^2ds \Big) \Big] \leq \frac{1}{5}\|M_1 - M_2 \|_{\BS^2}
		$$
	}
	Therefore, we have
	{\small
		$$
		\|F_t(V_1, M_1) - F_t(V_2, M_2)\|_{\BS^2} \leq  \frac{1}{5T} \BE \Big[ T \sup_{0\leq t \leq T} |V_1 -V_2|^2\Big]  + \frac{1}{5}\|M_1 - M_2\|_{\BS^2}
		= \frac{1}{5}  \Big( \|V_1 -V_2\|_{\BS^2} + \|M_1 - M_2\|_{\BS^2} \Big).
		$$}
	\endproof
	
	\subsection{Existence and Uniqueness of Solution}\label{sec:bsde_existunique}
	
	In this section, we will prove the existence and uniqueness of the solutions of the general BSDE. For BSDEs, we can prove the existences of solutions of a BSDE by proving the existence of the solutions for a corresponding fixed point problem, more details in \cite{cheridito2017bse}.
	
	\begin{thm}\label{thmbsde}
		If the BSDE \eqref{generalbsde} satisfies Assumption \ref{existuniqueass}, then the BSDE \eqref{generalbsde} admits a unique solutions $(V, Z, Z^I, Z^C, Z^\beta) \in \BS^2 \times \mathscr M$.
	\end{thm}
	
	\proof{Proof}
	By Proposition \ref{lipF}, the generator $F_t(V,M)$ satisfies 
	$$
	\|F_t(V_1, M_1) - F_t(V_2, M_2)\|_{\BS^2} < \frac{1}{5}\Big(\|V_1 - V_2\|_{\BS^2} + \| M_1 - M_2\|_{\BS^2}\Big).
	$$
	Since the generator satisfies the terminal condition and integrability condition, by the theorem 3.1 in \cite{cheridito2017bse}, the BSDE \eqref{generalbsde} admits a unique solution $(V,M) \in \BS^2 \times \mathscr M$.  ( Compared with the notation $F_t(k)(V,M)$ in \cite{cheridito2017bse}, we only need $ F_t(k)(V,M) : \equiv F_t(V,M)$ in our case.)
	
	Since $M\in \mathscr{M}$ is a martingale with respect to the filtration $(\calF_t)_{t \geq 0}$, by Theorem \ref{martdecom}, we can rewrite $M \in \mathscr M$ as
	$$
	M_T - M_t =  \int_t^T Z_s dW_s + \int_t^T Z^I_s d \varpi_t^I + \int_t^T Z^C_s d\varpi_t^C + \int_t^T Z^\beta_s d\tilde{J}_s + Y_T - Y_t.
	$$
	With the definition of the generator $F_t(V,M)$, the unique solution  $(V, Z, Z^I, Z^C, Z^\beta, Y) \in \BS^2 \times \mathscr M $ is represented as the form 
	$$
	V_t = \xi + \int_t^T f(\omega, s, V_{s-}, Z_s, Z^I_s, Z^C_s, Z^\beta_s) ds + \int_t^T Z_s dW_s + \int_t^T Z^I_s d \varpi_t^I + \int_t^T Z^C_s d\varpi_t^C + \int_t^T Z^\beta_s d\tilde{J}_s + Y_T - Y_t.
	$$
	Based on the form of the general BSDE \eqref{generalbsde} and the uniqueness of its solution, the orthogonal term $Y_T- Y_t \equiv 0$. Therefore, the unique solution of the general BSDE \eqref{generalbsde} is 
	$$
	V_t = \xi + \int_t^T f(\omega, s, V_{s-}, Z_s, Z^I_s, Z^C_s, Z^\beta_s) ds + \int_t^T Z_s dW_s + \int_t^T Z^I_s d \varpi_t^I + \int_t^T Z^C_s d\varpi_t^C + \int_t^T Z^\beta_s d\tilde{J}_s .
	$$
	
	\endproof

\bibliographystyle{spmpsci}
\bibliography{references_thesis}

\end{document}